\documentclass{handbook}

\usepackage{booktabs}
\usepackage{microtype}
\usepackage{mathtools}
\usepackage{bm}
\usepackage{bbm}
\usepackage{float}
\usepackage{wrapfig}
\usepackage{caption}
\usepackage{subcaption}
\usepackage{comment}
\usepackage{latexsym}
\usepackage{mathrsfs}
\usepackage{accents}
\usepackage{cases}
\usepackage[noabbrev]{cleveref}
\usepackage{algorithm}
\usepackage{algorithmic}
\usepackage{tabularx}
\usepackage{xcolor}

\newcommand{\D}{\mathcal{D}}
\newcommand{\A}{\mathcal{A}}
\newcommand{\B}{\mathcal{B}}

\newcommand{\pbm}{\mathbb{P}}
\newcommand{\E}{\mathbb{E}}
\newcommand{\W}{\mathcal{W}}
\newcommand{\disc}{D_{\omega}}
\newcommand{\gene}{G_{\theta}}
\newcommand{\gt}{g_{\theta}}
\newcommand{\go}{g_{\omega}}
\newcommand{\gtij}{g_{\theta}^{i,j}}
\newcommand{\goij}{g_{\omega}^{i,j}}

\newcommand{\gtb}{g_{\theta}^{\mathcal{B}}}
\newcommand{\gob}{g_{\omega}^{\mathcal{B}}}
\newcommand{\gtbk}{g_{\theta}^{I_k,J_k}}
\newcommand{\gobk}{g_{\omega}^{I_k,J_k}}
\newcommand{\sigt}{\Sigma_\theta}
\newcommand{\sigo}{\Sigma_\omega}
\newcommand{\params}{\theta,\omega}

\newcommand{\R}{\mathbb{R}}
\newcommand{\X}{\mathcal X}

\newcommand{\lrt}{\eta^\theta}
\newcommand{\lro}{\eta^\omega}

\newcommand\EE{\mathbb{E}}
\newcommand\RR{\mathbb{R}}
\newcommand\TT{\mathbb{T}}

\newcommand{\PP}{\mathbb{P}}

\newcommand{\ba}{\begin{array}}
\newcommand{\ea}{\end{array}}

\newcommand{\br}{\mathbb{R}}

\newcommand{\bp}{\mathbb{P}}

\newcommand{\XCal}{\mathcal{X}}

\newcommand{\ZCal}{\mathcal{Z}}

\newcommand{\bpm}{\begin{pmatrix}}
\newcommand{\epm}{\end{pmatrix}}

\newtheorem{thm}{Theorem}[chapter]

\newtheorem{prop}[thm]{Proposition}

\newtheorem{theorem}{Theorem}[chapter]
\newtheorem{definition}[theorem]{Definition}

\newtheorem{remark}[thm]{Remark}

\begin{document}

\startsubchapters

\chapter[GANs, MFGs and SDEs]{Generative Adversarial Network: Some Analytical  Perspectives}
\chapterauthor[1]{Haoyang Cao}
 \chapterauthor[2]{Xin Guo}
 \begin{affils}
   \chapteraffil[1]{The Alan Turing Institute, Email: hcao@turing.ac.uk}
   \chapteraffil[2]{University of California, Berkeley, Department of Industrial Engineering and Operations Research, Email: xinguo@berkeley.edu}
 \end{affils}

\begin{abstract}
    
    Ever since its debut, generative adversarial networks (GANs) have attracted tremendous amount of attention. Over the past years, different variations of GANs models have been developed and tailored to different applications in practice. Meanwhile, some issues regarding the performance and training of GANs have been noticed and investigated from various theoretical perspectives. This subchapter will start from an introduction of GANs from an analytical perspective, then move on to the training of GANs via SDE approximations and finally discuss some applications of GANs in computing high dimensional MFGs as well as tackling mathematical finance problems. 
\end{abstract}

\section{Introduction}

Generative adversarial networks (GANs), was introduced in 2014 to the machine learning community by \citep{Goodfellow2014}.  
The key idea behind GANs is to interpret the process of generative modeling as a competing game between two neural networks: a generator $G$ and a discriminator $D$. The generator attempts to fool the discriminator by converting random noise into sample data, while the discriminator tries to identify whether the input sample is fake or true.

Since its introduction,  GANs have enjoyed great empirical success, with a wide range of  applications especially in  image generation and natural language processing, including high resolution image generation \citep{denton2015deep,radford2015unsupervised}, image inpainting \citep{yeh2016semantic}, image super-resolution \citep{ledig2016others}, visual manipulation \citep{zhu2016generative}, text-to-image synthesis \citep{reed2016generative}, video generation \citep{vondrick2016generating}, semantic segmentation \citep{luc2016semantic}, and abstract reasoning diagram generation \citep{ghosh2016contextual}. 

Despite the empirical success of GANs, there are well recognized issues in GANs training, such as the vanishing gradient when the discriminator significantly outperforms the generator \citep{Arjovsky2017a}, the mode collapse which is believed to be linked with gradient exploding \citep{Salimans2016}, and the challenge of GANs convergence \citep{Barnett2018}. To improve the performance of GANs training, various approaches have been proposed for amelioration, including  different choices of network architectures, loss functions, and regularization. See for instance, a comprehensive survey on these techniques \citep{Wiatrak2019} and the references therein.
Meanwhile, there has been a growing research interest in the theoretical understanding of GANs training. \citep{Berard2020} proposes a novel visualization method for the GANs training process through the gradient vector field of loss functions.
In a deterministic GANs training framework, \citep{Mescheder2018} demonstrates that regularization improved  the convergence performance of GANs. \citep{Conforti2020} and \citep{Domingo-Enrich2020} analyze a generic zero-sum minimax game including that of GANs,
and connect the mixed Nash equilibrium  of the game with the invariant measure of Langevin dynamics. 

Recently, GANs have attracted attention in the mathematical finance community, largely due to 
the clear analogue between simulation of financial time series data and image generation, see for instance \citep{Wiese2019} and \citep{Wiese2020}. In response to the growing interests of GANs and its computational potential for high-dimensional control problems, stochastic games and backward-stochastic-differential equations, this note provides a gentle introduction of GANs from an analytical perspective, highlights 
some of the latest development of GANs training in the framework of  stochastic differential equations and reviews several representatives GANs applications in asset pricing and simulations of financial time series data.

Throughout this subchapter, the following notations will be adopted, unless otherwise specified.
\begin{itemize}
 \item The set of $k$ continuously differentiable functions over some domain $\mathcal X\subset\R^{d}$ is denoted by $\mathcal C^k(\mathcal X)$ for $k=0,1,2,\dots$; in particular when $k=0$, $\mathcal C^0(\mathcal X)=\mathcal C(\mathcal X)$ denotes the set of continuous functions.
 \item Let $p\geq1$. $L^p_{loc}(\R^d)$ denotes the set of functions $f$ defined on $\R^d$ such that for any compact subset $\X$, $\int_\X \|f(x)\|_p^pdx<\infty$.  
 \item Let $J=(J_1,\dots,J_d)$ be a $d$-tuple multi-index of order $|J|=\sum_{i=1}^dJ_i$. For a function ${f\in L^{1}_{loc}(\R^d)}$, its $J^{th}$-weak derivative ${D^Jf\in L^{1}_{loc}(\R^d)}$ is a function such that for any smooth and compactly supported test function $g$, \[\int_{\R^d}D^Jf(x)g(x)dx=(-1)^{|J|}\int_{\R^d}f(x)\nabla^Jg(x)dx.\] 
 \item The Sobolev space $W^{k,p}_{loc}(\R^d)$ is a set of functions $f$ on $\R^d$ such that for any $d$-tuple multi-index $J$ with $|J|\leq k$, $D^Jf\in L^p_{loc}(\R^d)$. 
\end{itemize}

\section{Basics of GANs: an analytical view}\label{sec: review}
\paragraph{GANs as generative models.}
GANs fall into the category of generative models.
 The procedure of generative modeling is to approximate an unknown probability distribution $\bp_r$ by constructing a class of suitable parametrized probability distributions $\bp_\theta$. That is, given a latent space $\ZCal$ and a sample space $\XCal$, define a latent variable $Z\in\ZCal$ with a fixed probability distribution $\PP_z$ and a family of functions $G_\theta: \ZCal\to\XCal$ parametrized by $\theta$. Then $\bp_\theta$ is defined as the probability distribution of $G_\theta(Z)$, i.e., $Law (G_\theta(Z))$. 

To distinguish from other generative models, GANs consist of two competing components: a generator $G$ and a discriminator $D$. In particular, the generator $G$ is implemented using a neural network (NN), i.e., function approximators via specific graph structures and network architectures, and it is denoted by $G=G_{\theta}$ as a parametrized function. Meanwhile, another neural network for the discriminator $D$ assigns a score between $0$ to $1$ to an input sample, either from the true distribution ${\mathbb P}_r$ or the approximated distribution ${\mathbb P}_{\theta} = Law(G_\theta(Z))$; denote the parametrized $D$ as $D_\omega$. A higher score from the discriminator $D$ would indicate that the sample is more likely to be from the true distribution. GANs are trained by optimizing $G$ and $D$ iteratively until $D$ can no longer distinguish between samples from $\bp_r$ and those from $\bp_\theta$. 

\paragraph{GANs as minimax games.}
Mathematically, GANs are minimax games as
\begin{equation} \label{gan-obj}
\begin{aligned}
	\min_G \max_D& \left\{\EE_{X \sim \bp_r} [\log D(X)]+ \EE_{Z \sim \bp_z} [\log (1 - D(G(Z)))]\right\}.
	\end{aligned}
\end{equation}
In particular, fixing $G$ and optimizing for $D$ in \eqref{gan-obj}, the optimal discriminator would be 
\[D^*_G(x) = \frac{p_r(x)}{p_r(x) + p_\theta(x)},\]
where $p_r$ and $p_\theta$ are density functions of $\bp_r$ and $\bp_\theta$ respectively. Plugging the above $D^*_G$ back to Equation \eqref{gan-obj}, the following equation holds,
\begin{align*}
&
\begin{aligned}
 \min_G &\left\{\EE_{X \sim \bp_r}\left[\log \frac{p_r(X)}{p_r(X) + p_\theta(X)}\right] + \EE_{Y \sim \bp_\theta} \left[\log \frac{p_\theta(Y)}{p_r(Y) + p_\theta(Y)}\right]\right\} 
\end{aligned}
	\\
	& \hspace{10pt}= -\log 4 + 2 JS(\bp_r, \bp_\theta). \label{gan-obj2}
\end{align*}

That is to say, training of GANs with Equation \eqref{gan-obj} being the objective is equivalent to minimizing Jensen-Shannon (JS) divergence between $\bp_r$ and $\bp_\theta$. In other words, through optimization over discriminators, GANs are essentially minimizing proper divergences between the true distribution and the generated distribution over some sample space $\XCal$.

\paragraph{GANs and optimal transport.}
This view of GANs as an optimization problem with an appropriate divergence function has been instrumental for addressing the instability of GANs training. Variants of GANs with different divergences have been proposed to improve the performance of GANs. For instance, \citep{nowozin2016f} and \citep{nock2017f} extend the JS divergence in \citep{Goodfellow2014} to a broader class of f-divergence. This extension provides the flexibility of choosing various $f$ functions for the loss function in GANs training. \citep{srivastava2019bregmn} explores scaled Bregman divergence to resolve the issue of support mismatch between $\bp_r$ and $\bp_\theta$ in the use of f-divergence and Bregman divergence. This is achieved through introducing a noisy base measure $\mu$ such that $\mu$ is a mixture of $\bp_r$ and $\bp_\theta$ convolved with some Gaussian distributions. \citep{Arjovsky2017} adopts Wasserstein-1 distance that enjoys higher smoothness with respect to the model parameters and consequently leads to a much more stable training of GANs. \citep{guo2017relaxed} proposes relaxed Wasserstein divergence by generalizing Wasserstein-1 distance with Bregman cost functions to first bypass the restriction on data information geometry in WGAN and achieve faster training. \citep{salimans2018improving} and \citep{sanjabi2018convergence} utilize the Sinkhorn loss instead of optimal transport type of loss by interpolating with energy distance and adding entropy regularization. This can significantly reduce the computational burden of optimal transport cost and increase stability of training.

The flexibility of choosing appropriate divergence functions, especially the development of  WGANs, leads to the natural connection between GANs and optimal transport problems, established in \citep{cao2020connecting}, which identifies sufficient conditions to recast GANs in the framework of optimal transport.

The idea behind this link is intuitive: GANs as generative models are minimax games with the goal to minimize the ``error'' of the generated sample data against the true sample data; this error is measured under appropriate divergence functions between the true distribution and the generated distribution. Now if this error is viewed as a cost of transporting/fitting the generated distribution into the true distribution, GANs become  optimal transport problems.

Indeed, this connection is explicit in the case of WGANs, via the Kantonovich duality  
\begin{theorem}
Suppose that $\PP_r\in L^1(\XCal)$ and $G\in L^1(\PP_z)$ where
\[L^1(\PP_z)=\biggl\{f:\ZCal\to\br:\int_\ZCal |f(z)|\PP_z(dz)<\infty\biggl\}.\]
WGAN is an optimal transport  problem between $Law(G(Z))$ and $\PP_r$.
\end{theorem}
As seen in \citep{cao2020connecting}, this connection  goes beyond the framework of WGANs.
Indeed, take any Polish space $\XCal$ with metric $l$, then $\XCal\times\XCal$ is also a Polish space with metric $l'$. Denote $\mathcal{P}(\XCal)$ as the set of all probability distributions over the sample space $\XCal$, define a generic divergence function
\[W:\mathcal{P}(\XCal)\times\mathcal{P}(\XCal)\mapsto\mathbb{R}^{+},\]
and take a class of GANs with this divergence $W$. 
 If $W$ can be written as an appropriate optimal cost $W_c$  and if such an optimal transport problem has a duality representation, then GANs model is an transport problem: 
the discriminator locates the best coupling among $\Pi_G$ under a given $G$, and  the generator refines the set of possible couplings $\Pi_G$ to minimize the divergence.

There are earlier studies connecting GANs and optimal transport problems, by different approaches and from different perspectives. For instance, 
\citep{salimans2018improving} defines a novel divergence called the minibatch energy distance, based on solutions of three associated 
optimal transport problems. This new divergence is then used to replace the JS divergence for the vanilla GANs. Note that this minibatch energy distance itself is not an optimal transport cost. In \citep{lei2019geometric}, a geometric interpretation of Wasserstein GANs (WGANs) from the perspective of optimal transport is provided: the latent random variable from the latent space is mapped to the sample space via an optimal mass transport so that the resulted distribution can minimize its Wasserstein distance against the true distribution. 

\paragraph{GANs and MFGs.}\label{sec: gans-mfgs}
In addition to this relation between GANs and optimal transport, \citep{cao2020connecting} further associates  GANs with mean-field games (MFGs), and design a new algorithm for computing MFGs. This connection between MFGs and GANs can be seen conceptually through the following Table \ref{tab: gan-mfg}.

\begin{table*}[!ht]
\centering
\caption{A first link between GANs and MFGs}
\begin{tabular*}{1.0\textwidth}{>{\bfseries}p{0.24\textwidth} p{0.24\textwidth} p{0.44\textwidth}}
\toprule
& {\textbf{GANs}} & {\textbf{MFGs}} \\
\toprule
Generator G & NN for approximating the map $G:\mathcal Z\mapsto \XCal$ & NN for solving HJB \\
\midrule
Characterization of $\mathbb{P}_r$ & Sample data & FP equation for consistency \\
\midrule
Discriminator D & NN measuring divergence between $\mathbb{P}_\theta$ and $\mathbb{P}_r$ & NN for measuring differential residual from the FP equation\\
\bottomrule
\end{tabular*}
\label{tab: gan-mfg}
\end{table*}

Evidently, there is more than one way to establish this connection between MFGs and GANs. Alternatively, one can switch the roles of the generator and discriminator and view the mean-field term as a generator and the value function as a discriminator.

For certain classes of MFGs,  such an interpretation of MFGs as GANs may be explicit.
For instance, take the  class of periodic MFGs from \citep{Cirant2018} on flat torus $\TT^d$ and a finite time horizon $[0,T]$. Such an MFG minimizes the following cost,
\begin{equation}\label{eq:dyn-cost}
 J_m(t,\alpha) = \EE\left[\int_t^T L(X^\alpha_t, \alpha(X^\alpha_t)) + f(X^\alpha_t, m(t, X^\alpha_t)) dt\right],\, t\in[0,T],
\end{equation}
where $X^\alpha = (X^\alpha_t)_t$ is a $d$-dimensional process with dynamics
\[
 dX^\alpha_t = \alpha(X^\alpha_t) dt + \sqrt{2 \epsilon} d W_t.
\]
Here $\alpha$ is a control policy, $L$ and $f$ constitute the running cost and $m(t,\cdot)$, for $t\in[0,T]$, denotes the probability density of $X^\alpha_t$ at time $t$. 

Now, consider the convex conjugate of the running cost $L$, namely,
\[
 H_0(x, p) = \sup_{\alpha \in \RR^d} \left\{ \alpha \cdot p - L(x, \alpha) \right\},
\]
and denote $F(x,m)=\int^mf(x,z)dz$. Then this class of MFGs can be characterized by the following coupled PDE system as illustrated in \citep{Cirant2018},
\begin{equation}
 \label{eq:dyn-hjb-fp}
 \begin{cases}
 &-\partial_s u-\epsilon\Delta_x u+H_0(x,\nabla_x u)=f(x,m),\\
 &\partial_sm-\epsilon\Delta_x m-div\left(m\nabla_pH_0(x,\nabla u)\right)=0, \\
 &m>0,\,m(0,\cdot)=m^0(\cdot),\,u(T,\cdot)=u^T(\cdot).
 \end{cases}
\end{equation}
Here the first equation is a Hamilton-Jacobi-Bellman (HJB) equation governing the value function and the second is a Fokker-Planck (FP) equation governing the evolution of the optimally controlled state process, with $m^0$ and $u^T$ the initial functions for $m(t,\cdot)$ and $u(t,\cdot)$, respectively. 

Note that this  system of equations \eqref{eq:dyn-hjb-fp} is equivalent to the following minimax game
\begin{equation}
 \label{eq:dyn-var-str}
 \inf_{u\in\mathcal C^2([0,T]\times\mathbb T^d)}\sup_{m\in\mathcal C^2([0,T]\times\mathbb T^d)}\Phi(m,u),
\end{equation}
where
\begin{equation*}
 \label{eq:dyn-functional}
 \begin{aligned}
 \Phi(m,u)&=\int_0^T\int_{\mathbb T^d}\left[m(-\partial_t u-\epsilon\Delta_x u)+mH_0(x,\nabla_x u)-F(x,m)\right] dxdt\\
 &\hspace{10pt}+\int_{\mathbb T^d}\left[m(T,x)u(T,x)-m^0(x)u(0,x)-m(x,T)u^T(x)\right]dx.
 \end{aligned}
\end{equation*}
Therefore, by \eqref{eq:dyn-var-str},  the connection between GANs and MFGs is transparent.

Having established the interpretation of MFGs as GANs, the immediate question to ask is whether GANs can be understood as MFGs. \citep{cao2020connecting} further shows that GANs can also be seen as MFGs, under the Pareto Optimality criterion. 
\begin{theorem}\label{thm: gan2mfg}
GANs in \citep{Goodfellow2014} are MFGs under the Pareto Optimality criterion, assuming that the latent variables $Z$ and true data $X$ are both i.i.d. sampled, respectively, with $\EE[|\log(D(X))|],\EE[|\log(1-D(G(Z)))|]<\infty$ for all possible $D$ and $G$. 
\end{theorem}
The above theorem shows that the theoretical framework of GANs in \citep{Goodfellow2014} can be seen as MFGs under Pareto Optimality criterion, where the generator network is an representative player of infinitely many identical players working in collaboration to defeat the discriminator. In practical training of GANs, however, only finitely many data points, i.e., $N$ latent variables $\{Z_i\}_{i=1}^N$ and $M$ samples from the unknown true distribution $\{X_j\}_{j=1}^M$, are available and therefore GANs in practice can be interpreted as $N$-player cooperative games with players being interchangeable and hence adopting the same strategy. Here, $Z_i\overset{i.i.d.}{\sim}\mathbb P_z$, $X_j\overset{i.i.d.}{\sim}\bp_r$ and $\{X_j\}_{j=1}^M\perp\{Z_i\}_{i=1}^N$. The state process for player $i$ is given by the feedforward process within its generator network network $G_i$, with the initial layer being $Z_i$ and the final layer being the generated sample $G_i(Z_i)$. Since the players are interchangeable and collaborating, a common generator network $G$ is adopted by all $N$ players to form a symmetric strategy profile $\mathbf S^G$ for the $N$-player game. These players face with a discriminator $D^{N,M}\in\mathcal D=\{D|D:\XCal\to[0,1]\}$ that favors the true samples $X_j$'s. In particular, the collective cost for the $N$ players of choosing a common generator $G$ is given by
\[J^{N,M}\left(\mathbf S^G;\{Z_i\}_{i=1}^N,\{X_j\}_{j=1}^M\right)=\max_{D\in\mathcal D}\frac{\sum_{i=1}^N\sum_{j=1}^M\log\left[D(X_j)\right] + \log\left[1-D(G(Z_i))\right]}{N\cdot M}\]
and $D^{N,M}$ is given by
\[\begin{aligned}
D^{N,M}&=D^{N,M}(\cdot;G,\{Z_i\}_{i=1}^N,\{X_j\}_{j=1}^M)\\
&=\arg\max_{D\in\mathcal D}\frac{\sum_{i=1}^N\sum_{j=1}^M\log\left[D(X_j)\right] + \log\left[1-D(G(Z_i))\right]}{N\cdot M}.
\end{aligned}\]
\begin{definition}[Pareto optimality]\label{defn: po}
A strategy profile $\mathbf S^{G^{N,*}}$ among all possible symmetric strategy profiles is said to be Pareto optimal if for any symmetric strategy profile $\mathbf{S}^G$,
\[J^{N,M}\left(\mathbf{S}^{G^{N,*}};\{Z_i\}_{i=1}^N,\{X_j\}_{j=1}^M\right)\leq J^{N,M}\left(\mathbf{S}^G;\{Z_i\}_{i=1}^N,\{X_j\}_{j=1}^M\right).\]
\end{definition}
Before characterizing $G^{N,*}$ and $D^{N,M}$, the cost $J^{N,M}$ using the empirical measures $\delta^M_r=\frac{1}{M}\sum_{j=1}^M\delta_{X_j}$ and $\delta^N_G=\frac{1}{N}\sum_{i=1}^N\delta_{G(Z_i)}$ can be rewritten as follows,
\[J^{N,M}\left(\mathbf S^G;\{Z_i\}_{i=1}^N,\{X_j\}_{j=1}^M\right)=\max_{D\in\mathcal D}\int_{\XCal}\log D(x)\delta^M_r(x)+\log[1-D(x)]\delta^N_G(x)dx.\]
Then $D^{N,M}$ and $G^{N,*}$ are naturally characterized by the two empirical distributions.
\begin{prop}
 Under a given $G$, a particular $D^{N,M}$ is given by
 \[
    D^{N,M}(x)=\begin{cases}
    \frac{\delta_r^M(x)}{\delta_r^M(x)+\delta_G^N(x)},\,x\in\{G(Z_1),\dots,G(Z_N),X_1,\dots,X_M\};\\
    \frac{1}{2},\,{\rm otherwise};
    \end{cases}
 \]
 in fact, for $x\not\in\{G(Z_1),\dots,G(Z_N),X_1,\dots,X_M\}$, $D^{N,M}\left(x\right)$ can take any value in $[0,1]$.
\end{prop} 
\begin{thm}
The set of possible $G^{N,*}$'s is given by
\[\mathcal G^{N,*}=\biggl\{G\in\mathcal G: \delta^N_G=\delta^M_r\biggl\},\]
provided that $\mathcal G^{N,*}\neq\emptyset$.
\end{thm}
The above results show that in practice training of GANs over finitely many samples, the generator can recover the empirical distribution of true samples at best. Moreover, the non-emptiness of $\mathcal G^{N,*}$ highly depends on the design of $G$ network architecture. This will be discussed in detail in the Section \ref{sec: gans-sdes}. Theoretically, however, $N$ and $M$ can be taken to infinity, leading the $N$-player cooperative games into MFGs with Pareto optimality criterion as stated in Theorem \ref{thm: gan2mfg}. Here the mean-field information is given by $G\#\bp_z=\lim_{N\to\infty}\delta^N_G$ and the convergence of $N$-player games to MFGs is guaranteed by the law of large numbers and the continuous mapping theorem.

\section{GANs Training}\label{sec: gans-sdes}
In the previous section, it has been pointed out that the success of GANs training depends on the design of network architecture. Apart from choosing a proper network architecture, there have been many practical methods to improve the performance of GANs training. This section is intended to provide mathematical explanation for these practical methods by analyzing GANs training via stochastic differential equation approximation. Before going into detail about GANs training, it is worth revisiting the objective of GANs. 
\paragraph{Equilibrium of GANs training.}
GANs are trained by optimizing $G$ and $D$ iteratively until $D$ can no longer distinguish between true samples and generated samples. Recall that $\gene$ denotes the generator parametrized by the neural network with the set of parameters $\theta\in\R^{d_\theta}$, and $\disc$ denotes the discriminator parametrized by the other neural network with the set of parameters $\omega\in\R^{d_\omega}$. Under a fixed network architecture, the parametrized version of GANs training is to find
\begin{equation} \label{gan-obj-param}
\begin{aligned}
	v_U^{GAN}&=\min_\theta \max_\omega L_{GAN}(\theta,\omega),\\
\text{where }L_{GAN}(\theta,\omega)&=\EE_{X \sim \bp_r} [\log D_\omega(X)]+ \EE_{Z \sim \bp_z} [\log (1 - D_\omega(G_\theta(Z)))].
	\end{aligned}
\end{equation}
\begin{remark}
From a game theory viewpoint, the objective in \eqref{gan-obj-param}, if attained, is in fact the upper value of the two-player zero-sum game of GANs. 
Meanwhile, the lower value of the game is given by the following maximin problem,
\begin{equation} \label{gan-obj-maximin}
\begin{aligned}
	v_L^{GAN}=\max_\omega\min_\theta L_{GAN}(\theta,\omega).
	\end{aligned}
\end{equation}

Clearly the following relation holds,
\begin{equation}
 \label{eq:dual-gap}
 v_L^{GAN}\leq v_U^{GAN}.
\end{equation}
Moreover, if there exists a pair of parameters $(\theta^*,\omega^*)$ such that both \eqref{gan-obj-param} and \eqref{gan-obj-maximin} are attained, then $(\theta^*,\omega^*)$ is a Nash equilibrium of this two-player zero-sum game. Indeed, if $L_{GAN}$ is convex in $\theta$ and concave in $\omega$, then there is no duality gap hence the equality in \eqref{eq:dual-gap} holds by the minimax theorem (see \citep{von1959theory} and \citep{sion1958general}). 

It is worth noting that conditions for such an equality in \eqref{eq:dual-gap} is usually not satisfied in many common GANs models, as observed in \citep{zhu2020deconstructing} and analyzed in 
\citep{guo2020optimal}.
\end{remark}
\paragraph{GANs training via SGD.}
As in most deep learning models, stochastic gradient descent (SGD) (or one of its variants) is a standard approach for solving the optimization problem in GANs training.
Accordingly, the evolution of parameters of $\theta$ and $\omega$ in \eqref{gan-obj-param} by SGD from current step $t$ to the next step $t+1$ is
\begin{equation}
\label{eq:evol}
\begin{aligned}
 &\omega_{t+1}=\omega_t+\alpha_d\nabla_{\omega} L_{GAN}(\theta_t,\omega_t),\\
 &\theta_{t+1}=\theta_t-\alpha_g\nabla_{\theta} L_{GAN}(\theta_t,\omega_{t+1}).
\end{aligned}
\end{equation}
Here the $\alpha_d$ and $\alpha_g$ denote the step sizes of updating the discriminator and the generator, respectively. Evolution \eqref{eq:evol} corresponds to the alternating updating scheme of the algorithm in \citep{Goodfellow2014} where at each iteration, the discriminator is updated before the generator. One of the main challenges for GANs training is the convergence of such an alternating SGD.

\paragraph{GANs training and SDEs approximation.}
GANs training is performed on a data set $\D=\{(z_i,x_j)\}_{1\leq i\leq N,\,1\leq j\leq M}$, where $\{z_i\}_{i=1}^N$ are sampled from $\pbm_z$ and $\{x_j\}_{j=1}^M$ are real image data following the unknown distribution $\pbm_r$. The objective of GANs is to solve the following minimax problem
\begin{equation}
    \label{eq: minimax-obj}
    \min_\theta\max_\omega\Phi(\theta,\omega),
\end{equation}
 for some cost function $\Phi$, with $\Phi$ of a separable form
 \begin{equation}
     \Phi(\params)=\frac{\sum_{i=1}^N\sum_{j=1}^MJ(\disc(x_j),\disc(\gene(z_i)))}{N\cdot M}.
 \end{equation}
 
When the stochastic gradient algorithm (SGA) is performed to  solve the minimax problem \eqref{eq: minimax-obj}, the full gradients of $\Phi$ with respect to $\theta$ and $\omega$, denoted by $g_\theta$ and $g_\omega$ respectively, are estimated over a mini-batch $\B$ of batch size $B$, denoted by $g_\theta^{\B}$ and $g_\omega^{\B}$. 

Let $\eta^\theta_t>0$ and $\eta^\omega_t>0$ be the learning rates at iteration $t=0,1,2,\dots$, for $\theta$ and $\omega$ respectively, then solving the minimax problem \eqref{eq: minimax-obj} with SGA and {\em alternating parameter update} implies descent of $\theta$ along $\gt$ and ascent of $\omega$ along $\go$ at each iteration; within each iteration, the minibatch gradient for $\theta$ and $\omega$ are calculated on different batches. In order to emphasize this difference, $\bar\B$ represents the minibatch for $\theta$ and $\B$ for that of $\omega$, with $\bar\B\overset{i.i.d.}{\sim}\B$. The one-step update can be written as follows.
\begin{equation}
    \label{eq: dt-gan-update}\tag{ALT}
    \left\{\begin{aligned}
      &\omega_{t+1}=\omega_t+\lro_t\gob(\theta_t,\omega_t),\\
      &\theta_{t+1}=\theta_t-\lrt_t\gt^{\bar\B}(\theta_t,\omega_{t+1}).
    \end{aligned}\right.
\end{equation}
Some practical training of GANs uses {\em simultaneous parameter update} between the discriminator and the generator, corresponding to a similar yet subtly different form
\begin{equation}
    \label{eq: sml-update}\tag{SML}
    \left\{\begin{aligned}
      &\omega_{t+1}=\omega_t+\lro_t\gob(\theta_t,\omega_t),\\
      &\theta_{t+1}=\theta_t-\lrt_t\gtb(\theta_t,\omega_{t}).
    \end{aligned}\right.
\end{equation}

For the ease of exposition, the learning rates are assumed to be constant $\lrt_t=\lro_t=\eta$, with $\eta$ viewed as the time interval between two consecutive parameter updates. In \citep{guo2020optimal}, the optimal (variable) learning rate for GANs training is studied under a stochastic control framework. 

Let $\gtij$ and $\goij$ denote $\nabla_\theta J(D_\omega(x_j),D_\omega(G_\theta(z_i)))$ and $\nabla_\omega J(D_\omega(x_j),D_\omega(G_\theta(z_i)))$, respectively, and define the following covariance matrices
\[\begin{aligned}
  &\sigt(\params)=\frac{\sum_i\sum_j[\gtij(\params)-\gt(\params)][\gtij(\params)-\gt(\params)]^T}{N\cdot M},\\
  &\sigo(\params)=\frac{\sum_i\sum_j[\goij(\params)-\go(\params)][\goij(\params)-\go(\params)]^T}{N\cdot M}, 
\end{aligned}\]
then as the batch size $B$ gets sufficiently large, the classical central limit theorem leads to the following approximation of \eqref{eq: dt-gan-update},
\begin{equation}\label{eq: gan-evol}
\left\{\begin{aligned}
  \omega_{t+1}&=\omega_t+\eta\gob(\theta_t,\omega_t)\approx \omega_t+\eta\go(\theta_t,\omega_t)+\frac{\eta}{\sqrt{B}}\sigo^{\frac{1}{2}}(\theta_t,\omega_t)Z^1_t,\\
  \theta_{t+1}&=\theta_t-\eta\gtb(\theta_{t},\omega_{t+1})\approx \theta_t-\eta\gt(\theta_t,\omega_{t+1})+\frac{\eta}{\sqrt{B}}\sigt^{\frac{1}{2}}(\theta_t,\omega_{t+1})Z^2_t,
\end{aligned}\right.
\end{equation}
with independent random variables $Z^1_t{\sim} N(0,I_{d_\omega})$ and $Z^2_t\sim N(0,I_{d_\theta})$, $t=0,1,2,\dots$.

If ignoring the difference between $t$ and $t+1$, then the approximation could be written in the following form
 \begin{equation}
    \label{eq: pre-dyn-approx}
    \begin{aligned}
    d\begin{pmatrix}\Theta_t\\\mathcal{W}_t\end{pmatrix}=\begin{pmatrix}-\gt(\Theta_t,\W_t)\\\go(\Theta_t,\W_t)\end{pmatrix}dt
    +\sqrt{2\beta^{-1}}\begin{pmatrix}\sigt(\Theta_t,\W_t)^{\frac{1}{2}}&0\\
    0&\sigo(\Theta_t,\W_t)^{\frac{1}{2}}\end{pmatrix}dW_t,
    \end{aligned}
\end{equation}
with $\beta=\frac{2B}{\eta}$ and $\{W_t\}_{t\geq0}$ be standard $(d_\theta+d_\omega)$-dimensional Brownian motion.
This would be the approximation for GANs training of \eqref{eq: sml-update}.

Taking the subtle difference between $t$ and $t+1$ into consideration and thus the interaction  between the generator and the discriminator, the approximation for the GANs training process of \eqref{eq: dt-gan-update} should be 
  \begin{equation}
    \label{eq: pre-gan-dyn}
    \begin{aligned}
    d\begin{pmatrix}\Theta_t\\\mathcal{W}_t\end{pmatrix}&=\biggl[\begin{pmatrix}-\gt(\Theta_t,\mathcal{W}_t)\\\go(\Theta_t,\W_t)\end{pmatrix}\\
    &\hspace{30pt}+\frac{\eta}{2}\begin{pmatrix}\nabla_\theta\gt(\Theta_t,\mathcal{W}_t)&-\nabla_\omega\gt(\Theta_t,\mathcal{W}_t)\\-\nabla_\theta\go(\Theta_t,\mathcal{W}_t)&-\nabla_\omega\go(\Theta_t,\mathcal{W}_t)\end{pmatrix}\begin{pmatrix}-\gt(\Theta_t,\mathcal{W}_t)\\\go(\Theta_t,\mathcal{W}_t)\end{pmatrix}\biggl]dt\\
    &\hspace{30pt}+\sqrt{2\beta^{-1}}\begin{pmatrix}\sigt(\Theta_t,\W_t)^{\frac{1}{2}}&0\\
    0&\sigo(\Theta_t,\W_t)^{\frac{1}{2}}\end{pmatrix}dW_t.
    \end{aligned}
\end{equation}
Equations \eqref{eq: pre-dyn-approx} and \eqref{eq: pre-gan-dyn} can be written in more compact forms
\begin{align}d\begin{pmatrix}\Theta_t\\\mathcal{W}_t\end{pmatrix}&=b_0(\Theta_t,\W_t)dt+\sigma(\Theta_t,\W_t)dW_t,\label{eq: dyn-approx}\tag{SML-SDE}\\ d\begin{pmatrix}\Theta_t\\\mathcal{W}_t\end{pmatrix}&=b(\Theta_t,\W_t)dt+\sigma(\Theta_t,\W_t)dW_t.\label{eq: gan-dyn}\tag{ALT-SDE}\end{align}
where $b(\params)=b_0(\params)+\eta b_1(\params)$, with
\begin{align}
    b_0(\params)&=\begin{pmatrix}-\gt(\params)\\\go(\params)\end{pmatrix},\\
    b_1(\params)&=\frac{1}{2}\begin{pmatrix}\nabla_\theta\gt(\params)&-\nabla_\omega\gt(\params)\\-\nabla_\theta\go(\params)&-\nabla_\omega\go(\params)\end{pmatrix}\begin{pmatrix}-\gt(\params)\\\go(\params)\end{pmatrix}\nonumber\\
    &=-\frac{1}{2}\nabla b_0(\params)b_0(\params)-\begin{pmatrix}\nabla_\omega\gt(\params)\go(\params)\\0\end{pmatrix},\label{eq: additional-term}\\
\text{and}\hspace{5pt}
    \sigma(\params)&=\sqrt{2\beta^{-1}}\begin{pmatrix}\sigt(\Theta_t,\W_t)^{\frac{1}{2}}&0\\
    0&\sigo(\Theta_t,\W_t)^{\frac{1}{2}}\end{pmatrix}.\label{eq: volatility}
\end{align}
Note the term $-\frac{\eta}{2}\begin{pmatrix}\nabla_\omega\gt(\params)\go(\params)\\0\end{pmatrix}$ for \eqref{eq: gan-dyn}, which highlights the interaction between the generator and the discriminator in GANs training process. 

In \citep{cao2020approximation}, it is shown that these coupled SDEs are indeed the continuous-time approximations of GANs training processes, with precise error bound analysis, where the approximations are under the notion of weak approximation as in \citep{Li2019}. 

\begin{thm}\label{thm: gan-approx}
Fix an arbitrary time horizon $\mathcal T>0$ and take the learning rate $\eta\in(0,1\wedge \mathcal T)$ and the number of iterations $N=\left\lfloor\frac{\mathcal T}{\eta}\right\rfloor$. Suppose that
\begin{enumerate}
    \item[1.a] $\goij$ is twice continuously differentiable, and $\gtij$ and $\goij$ are Lipschitz, for any $i=1,\dots,N$ and $j=1,\dots,M$;
    \item[1.b] $\Phi$ is of $\mathcal{C}^3(\R^{d_\theta+d_\omega})$, $\Phi\in W^{4,1}_{loc}(\R^{d_\theta+d_\omega})$, and for any multi-index $J=(J_1,\dots,J_{d_\theta+d_\omega})$ with $|J|=\sum_{i=1}^{d_\theta+d\omega}J_i\leq 4$, there exist $k_1,k_2\in\mathbb N$ such that
    \[|D^J\Phi(\theta,\omega)|\leq k_1\left(1+\left\|\begin{pmatrix}\theta\\\omega\end{pmatrix}\right\|_2^{2k_2}\right)\]
    for $\theta\in\R^{d_\theta}$, $\omega\in\R^{d_\omega}$ almost everywhere; 
    \item[1.c] $(\nabla_\theta\gt)\gt$, $(\nabla_\omega\gt)\go$, $(\nabla_\theta\go)\gt$ and $(\nabla_\omega\go)\go$ are all Lipschitz.
\end{enumerate}
Then, given any initialization  $\theta_0=\theta$ and $\omega_0=\omega$, for any test function $f\in\mathcal C^3(\R^{d_\theta+d_\omega})$ such that for any multi-index $J$ with $|J|\leq 3$ there exist $k_1, k_2\in\mathbb N$ satisfying 
\[|\nabla^Jf(\theta,\omega)|\leq k_1\left(1+\left\|\begin{pmatrix}\theta\\\omega\end{pmatrix}\right\|_2^{2k_2}\right),\]
 the following weak approximation holds
\begin{equation}
\label{eq: alt-error}
\max_{t=1,\dots,N}\left|\E f(\theta_t,\omega_t)-\E f(\Theta_{t\eta},\mathcal W_{t\eta})\right|\leq C\eta^2
\end{equation}
for constant $C\geq0$, where $(\theta_t,\omega_t)$ and $(\Theta_{t\eta},\mathcal W_{t\eta})$ are given by \eqref{eq: dt-gan-update} and \eqref{eq: gan-dyn}, respectively.
\end{thm}

\begin{thm}
\label{thm: sml-sde}
Fix an arbitrary time horizon $\mathcal T>0$, take the learning rate $\eta\in(0,1\wedge \mathcal T)$ and the number of iterations $N=\left\lfloor\frac{\mathcal T}{\eta}\right\rfloor$. Suppose
\begin{enumerate}
    \item[2.a] $\Phi(\theta,\omega)$ is continuously differentiable, $\Phi\in W^{3,1}_{loc}(\R^{d_\theta+d_\omega})$ and for any multi-index $J=(J_1,\dots,J_{d_\theta+d_\omega})$ with $|J|=\sum_{i=1}^{d_\theta+d\omega}J_i\leq 3$, there exist $k_1,k_2\in\mathbb N$ such that  $D^J\Phi$ satisfies
    \[|D^J\Phi(\theta,\omega)|\leq k_1\left(1+\left\|\begin{pmatrix}\theta\\\omega\end{pmatrix}\right\|_2^{2k_2}\right)\]
    for $\theta\in\R^{d_\theta}$, $\omega\in\R^{d_\omega}$ almost everywhere; 
    \item[2.b] $\gtij$ and $\goij$ are Lipschitz for any $i=1,\dots, N$ and $j=1,\dots,M$.
\end{enumerate}
Then, given any initialization $\theta_0=\theta$ and $\omega_0=\omega$, for any test function $f\in\mathcal C^2(\R^{d_\theta+d_\omega})$ such that for any multi-index $J$ with $|J|\leq 2$ there exist $k_1, k_2\in\mathbb N$ satisfying 
\[|\nabla^Jf(\theta,\omega)|\leq k_1\left(1+\left\|\begin{pmatrix}\theta\\\omega\end{pmatrix}\right\|_2^{2k_2}\right),\]
 then the following weak approximation holds
\begin{equation}\label{eq: sml-error}
\max_{t=1,\dots,N}\left|\E f(\theta_t,\omega_t)-\E f(\Theta_{t\eta},\mathcal W_{t\eta})\right|\leq C\eta
\end{equation} for constant $C\geq0$, where $(\theta_t,\omega_t)$ and $(\Theta_{t\eta},\mathcal W_{t\eta})$ are given by \eqref{eq: sml-update} and \eqref{eq: dyn-approx}, respectively.
\end{thm}
The above theorems from \citep{cao2020approximation} make it possible to analyze the convergence of GANs training via the invariant measure of the SDEs. 

\paragraph{Convergence of GANs training via invariant measure of SDEs.}

The invariant measure here in the context of GANs training can be interpreted in the following sense. First of all, the invariant measure $\mu^*$  describes the joint probability distribution of the generator and discriminator parameters $(\Theta^*, \W^*)$ in equilibrium. For instance, if the training process converges to the unique minimax point $(\theta^*,\omega^*)$ for $\min_{\theta}\max_{\omega}\Phi(\params)$, the invariant measure is the Dirac mass at $(\theta^*,\omega^*)$. Having the distribution of $(\theta^*,\omega^*)$, especially the marginal distribution of $\Theta^*$, helps to characterize the probability distribution of the generated samples, $Law(G_{\Theta^*}(Z))$, and this distribution is in particular useful in the evaluation of GANs performance via metrics such as inception score and Fr\`echet inception distance. (See \citep{Salimans2016, Heusel2017} for more details on these metrics). Besides, from a game perspective, the pair of conditional laws $(Law(\Theta^*|\W^*),Law(\W^*|\Theta^*))$ can be seen as the mixed strategies adopted by the generator and discriminator in equilibrium, respectively. 

\begin{theorem}
\label{thm: inv-msr}
Assume the following conditions hold for  \eqref{eq: gan-dyn}. 
\begin{enumerate}
    \item[3.a] both $b$ and $\sigma$ are bounded and smooth and have bounded derivatives of any order;
    \item[3.b] there exist some positive real numbers $r$ and $M_0$ such that for any $\begin{pmatrix}\theta\ \ \omega\end{pmatrix}^T\in\R^{d_\theta+d_\omega}$,
    \[\begin{pmatrix}\theta\ \ \omega\end{pmatrix} b(\params)\leq -r\left\|\begin{pmatrix}\theta\\\omega\end{pmatrix}\right\|_2,\text{ if }\left\|\begin{pmatrix}\theta\\\omega\end{pmatrix}\right\|_2\geq M_0;\]
    \item[3.c] $\A$ is uniformly elliptic, i.e., there exists $l>0$ such that for any $\begin{pmatrix}\theta\\\omega\end{pmatrix},\begin{pmatrix}\theta'\\\omega'\end{pmatrix}\in\R^{d_\theta+d_\omega}$,
    \[\begin{pmatrix}\theta'\ \ \omega'\end{pmatrix}^T\sigma(\params)\sigma(\params)^T\begin{pmatrix}\theta'\\\omega'\end{pmatrix}\geq l\left\|\begin{pmatrix}\theta'\\\omega'\end{pmatrix}\right\|_2^2,\] 
\end{enumerate}
then \eqref{eq: gan-dyn}  admits a unique invariant measure $\mu^*$ with an exponential convergence rate. 

Similar results hold for the invariant measure of  \eqref{eq: dyn-approx} with $b$ replaced by $b_0$. 
\end{theorem}

The assumptions 1.a-1.c, 2.a-2.b and 3.a for the regularity conditions of the drift, the volatility, and the derivatives of loss function $\Phi$, are more than  mathematical convenience. They are essential constraints on the growth of the loss function with respect to the model parameters, necessary for avoiding the explosive gradient encountered in the training of GANs. Moreover, these conditions put restrictions on the gradients of the objective functions with respect to the parameters. By the chain rule, it requires both careful choices of network structures as well as particular forms of the loss function $\Phi$.

\paragraph{Dynamics of training loss and FDR.}
To have a more quantifiable characteristic of the convergence of GANs training, the analysis of the training loss dynamics reveals a fluctuation-dissipation relation (FDR) for the GANs training.

\begin{theorem}
\label{thm: fdr}
Assume the existence of an invariant measure $\mu^*$ for \eqref{eq: gan-dyn}, then 
\begin{equation}
    \label{eq: fdr}\tag{FDR1}
    \begin{aligned}
      &\E_{\mu^*}\biggl[\|\nabla_\theta\Phi(\Theta^*,\W^*)\|_2^2-\|\nabla_\omega\Phi(\Theta^*,\W^*)\|_2^2\biggl]=\beta^{-1}\E_{\mu^*}\biggl[Tr\biggl(\sigt(\Theta^*,\W^*)\nabla_\theta^2\Phi(\Theta^*,\W^*)\\
      &\hspace{15pt}+\sigo(\Theta^*,\W^*)\nabla_\omega^2\Phi(\Theta^*,\W^*)\biggl)\biggl]-\frac{\eta}{2}\E_{\mu^*}\biggl[\nabla_\theta\Phi(\Theta^*,\W^*)^T\nabla_\theta^2\Phi(\Theta^*,\W^*)\nabla_\theta\Phi(\Theta^*,\W^*)\\
      &\hspace{15pt}+\nabla_\omega\Phi(\Theta^*,\W^*)^T\nabla_\omega^2\Phi(\Theta^*,\W^*)\nabla_\omega\Phi(\Theta^*,\W^*)\biggl].
    \end{aligned}
\end{equation}
The corresponding FDR for the simultaneous update case of  \eqref{eq: dyn-approx} is
\begin{equation*}
    \label{eq: fdr-sml}
    \begin{aligned}
    &\E_{\mu^*}\biggl[\|\nabla_\theta\Phi(\Theta^*,\W^*)\|_2^2-\|\nabla_\omega\Phi(\Theta^*,\W^*)\|_2^2\biggl]=\\
   &\hspace{20pt}\beta^{-1}
   \E_{\mu^*}\biggl[Tr\biggl(\sigt(\Theta^*,\W^*)\nabla_\theta^2\Phi(\Theta^*,\W^*)+\sigo(\Theta^*,\W^*)\nabla_\omega^2\Phi(\Theta^*,\W^*)\biggl)\biggl].
   \end{aligned}
\end{equation*}
\end{theorem}
Note that this FDR relation for GANs training is analogous to  that for stochastic gradient descent algorithm on a pure minimization problem in \citep{Yaida2019} and \citep{Liu2019b}.
This FDR relation in GANs reveals the crucial difference between GANs training of discriminator  and generator networks versus training of two independent neural networks. It connects the microscopic fluctuation from the noise of SGA with the macroscopic dissipation phenomena related to the loss function. In particular, the quantity $Tr(\sigt\nabla^2_\theta\Phi+\sigo\nabla^2_\omega\Phi)$  links the covariance matrices $\sigt$ and $\sigo$ from SGAs with the loss landscape of $\Phi$, and reveals the trade-off of the loss landscape between the generator and the discriminator. 

Alternatively, the evolution of the squared norm of the parameters leads to a different type of FDR that will be practically useful for learning rate scheduling. 
\begin{theorem}
\label{thm: fdr2}
Assume the existence of an invariant measure $\mu^*$ for \eqref{eq: dyn-approx}, then
\begin{equation}
    \label{eq: fdr2}\tag{FDR2}
   \E_{\mu^*}\biggl[\Theta^{*,T}\nabla_\theta\Phi(\Theta^*,\W^*)-\W^{*,T}\nabla_\omega\Phi(\Theta^*,\W^*)\biggl]=\beta^{-1}\E_{\mu^*}\biggl[Tr(\sigt(\Theta^*,\W^*)+\sigo(\Theta^*,\W^*))\biggl]
\end{equation}
\end{theorem}

\paragraph{Scheduling of learning rate.}
Notice that the quantities in \eqref{eq: fdr2}, including the parameters $(\params)$ and first-order derivatives of the  loss function $\gt$, $\go$, $\gtij$ and $\goij$, are computationally inexpensive. Therefore, \eqref{eq: fdr2}
enables customized scheduling of learning rate, 
instead of  predetermined scheduling ones such as Adam or RMSprop optimizer.

For instance, recall that $\gtb$ and $\gob$ are
respectively unbiased estimators for $\gt$ and $\go$, and 
\begin{align*}
&\hat\Sigma_\theta(\params)=\frac{\sum_{k=1}^B[\gtbk(\params)-\gtb(\params)][\gtbk(\params)-\gtb(\params)]^T}{B-1},\\
&\hat\Sigma_\omega(\params)=\frac{\sum_{k=1}^B[\gobk(\params)-\gob(\params)][\gobk(\params)-\gob(\params)]^T}{B-1}
\end{align*}
are respectively unbiased estimators of $\sigt(\params)$ and $\sigo(\params)$. Now in order  to improve GANs training  with the simultaneous update, one can introduce two tunable parameters $\epsilon>0$ and $\delta>0$ to have the following scheduling:
\begin{center}
if $\left|\frac{\Theta^T\gtb(\Theta_t,\W_t)-\W_t^T\gob(\Theta_t,\W_t)}{\beta^{-1}Tr(\hat\Sigma_\theta(\Theta_t,\W_t)+\hat\Sigma_\omega(\Theta_t,\W_t))}-1\right|<\epsilon$, then update $\eta$ by $(1-\delta)\eta$.
\end{center}

\section{Applications of GANs}\label{sec: gans-mf}
\subsection{Computing MFGs via GANs}
Bases on the conceptual connection between GANs and MFGs, \citep{cao2020connecting} proposes a new computational approach for MFGs, using two neural networks in an adversarial way, summarized in Algorithm \ref{alg:mfgan-dyn}, in which 
\begin{itemize}
 \item $u_\theta$ being the NN approximation of the unknown value function $u$ for the HJB equation, 
 \item $m_\omega$ being the NN approximation for the unknown mean information function $m$.
\end{itemize}
Note that Algorithm \ref{alg:mfgan-dyn} can be adapted for broader classes of dynamical systems with variational structures. Such GANs structures are exploited in \citep{Yang2018} and \citep{Yang2018a} to synthesize complex systems governed by physical laws. 
 
\begin{algorithm}
\caption{MFGANs}
\label{alg:mfgan-dyn}
\begin{algorithmic}
\STATE{At $k=0$, initialize $\theta$ and $\omega$. Let $N_{\theta}$ and $N_\omega$ be the number of training steps of the inner-loops and $K$ be that of the outer-loop. Let $\beta_i>0$, $i=1,2$.}
\FOR{$k\in\{0,\dots, K-1\}$}
 \STATE{Let $m=0$, $n=0$.}
 \STATE{Sample $\{(s_i,x_i)\}_{i=1}^{B_d}$ on $[0,T]\times \mathbb R^d$ according to a predetermined distribution $p_{prior}$, where $B_d$ denotes the number of training samples for updating loss related to FP residual. }
 \STATE{Let $\hat L_{D}(\theta, \omega)=\hat L_{FP}(\theta,\omega)+\beta_{D} \hat L_{init}(\omega)$,
 with 
 \[
 \begin{aligned}
 \hat L_{FP}&=\frac{1}{B_d}\biggl\{\sum_{i=1}^{B_d}\biggl[\partial_sm_\omega(s_i,x_i)+div\left[m_\omega(s_i,x_i)b(s_i,x_i,m(s_i,x_i),\alpha^*_{\theta,\omega}(s_i,x_i))\right]\\
 &\hspace{60pt}-\frac{\sigma^2}{2}\Delta_xm_\omega(s_i,x_i)\biggl]^2\biggl\},\\
 \hat L_{init}&=\frac{\sum_{i=1}^{B_d}\left[m_\omega(0,x_i)-m^0(x_i)\right]^2}{B_d},
 \end{aligned}\]
 where $m^0$ is a known density function for the initial distribution of the states and $\beta_{D}>0$ is the weight for the penalty on the initial condition of $m$.}
 \FOR{$m\in\{0,\dots, N_\omega-1\}$}
 \STATE{$\omega\leftarrow w-\alpha_d\nabla_\omega\hat L_{D}$ with learning rate $\alpha_d$.}
 \STATE{Increase $m$.}
 \ENDFOR
 \STATE{Sample $\{(s_j,x_j)\}_{j=1}^{B_g}$ on $[0,T]\times \mathbb R$ according to a predetermined distribution $p_{prior}$, where $B_g$ denotes the number of training samples for updating loss related to HJB residual. }
 \STATE{Let $\hat L_{G}(\theta,\omega)=\hat L_{HJB}(\theta,\omega)+\beta_{G}\hat L_{term}(\theta)$,
 with
 \[\begin{aligned}
 \hat L_{HJB}&=\frac{1}{B_g}\biggl\{\sum_{j=1}^{B_g}\left[\partial_su_\theta(s_j,x_j)+\frac{\sigma^2}{2}\Delta_xu_\theta(s_j,x_j)+H_\omega\left(s_j,x_j,\nabla_xu_\theta(s_j,x_j)\right)\right]^2\biggl\},\\
 \hat L_{term}&=\frac{\sum_{j=1}^{B_g}u_\theta(T,x_j)^2}{B_g},
 \end{aligned}
 \]
 where $\beta_{G}>0$ is the weight for the penalty on the terminal condition of $u$.
 }
 \FOR{$n\in\{0,\dots,N_\theta-1\}$}
 \STATE{$\theta\leftarrow\theta-\alpha_g\nabla_\theta\hat L_{G}$ with learning rate $\alpha_g$.}
 \STATE{Increase $n$}
 \ENDFOR
 \STATE{Increase $k$.}
\ENDFOR
\STATE Return $\theta$, $\omega$
\end{algorithmic}
\end{algorithm}

To test the performance of Algorithm \ref{alg:mfgan-dyn}, a class of ergodic MFGs with the following payoff function are considered,
\begin{equation}
 \label{eq:cost}
 \hat J_m(\alpha)=\liminf_{T\to\infty}\frac{1}{T}\EE\left[\int_t^T L(X^\alpha_t, \alpha(X^\alpha_t)) + f(X^\alpha_t, m(X^\alpha_t)) dt\right],
\end{equation}
subject to $dX^\alpha_t=\alpha(X^\alpha_t) dt+dW_t$, with the cost of control and running cost given by
\[
\begin{aligned}
&L(x,\alpha) = \frac{1}{2} |\alpha|^2 + 2 \pi^2 \left[ - \sum_{i=1}^d \sin(2 \pi x_i) + \sum_{i=1}^d |\cos(2 \pi x_i)|^2 \right]- 2\sum_{i=1}^d \sin(2 \pi x_i),\\
&f(x, m) = \ln(m).
\end{aligned}\]
In this class of mean-field-games, the associated HJB equation and FP equation are
\begin{equation}\label{eq:hjb-fp}
 \begin{cases}
 -\epsilon\Delta u+H_0(x,\nabla u)=f(x,m)+\bar H,\\
 -\epsilon\Delta m-div\left(m\nabla_pH_0(x,\nabla u)\right)=0,\\
 \int_{\mathbb T^d}u(x)dx=0;\,m>0,\, \int_{\mathbb T^d}m(x)dx=1,
 \end{cases}
\end{equation}
where the convex conjugate $H_0$ is given by
$H_0(x,p)=\sup_{\alpha}\{\alpha\cdot p-\frac{1}{2}|\alpha|^2\}-\tilde f(x)$.
Here, the periodic value function $u$, the periodic density function $m$, and the unknown $\bar H$ can be explicitly derived. Indeed, assuming the existence of a smooth solution $(m,u,\bar H)$, $m$ in the second equation in \eqref{eq:hjb-fp} can be written as $m(x) = \frac{e^{2u(x)}}{\int_{\mathbb T^d} e^{2u(x')}dx'}$. Hence the solution to \eqref{eq:hjb-fp} is given by 
$u(x) = \sum_{i=1}^d \sin(2 \pi x_i)$
and
$\bar H = \ln\left(\int_{\mathbb T^d} e^{2 \sum_{i=1}^d \sin(2 \pi x_i)} d x\right)$.
The optimal control policy is also explicitly given by
\[\begin{aligned}
\alpha^*&=\arg\max_{\alpha}\{\nabla_xu\cdot \alpha -L(x,\alpha)\}\\
&=\nabla_xu=2\pi\begin{pmatrix*}\cos(2\pi x_1)&\dots&\cos(2\pi x_d)\end{pmatrix*}\in\RR^d.
\end{aligned}\]

\begin{figure}[!ht]
 \centering
 \begin{minipage}[t]{0.4\columnwidth}
 \centering
 \includegraphics[width=\textwidth]{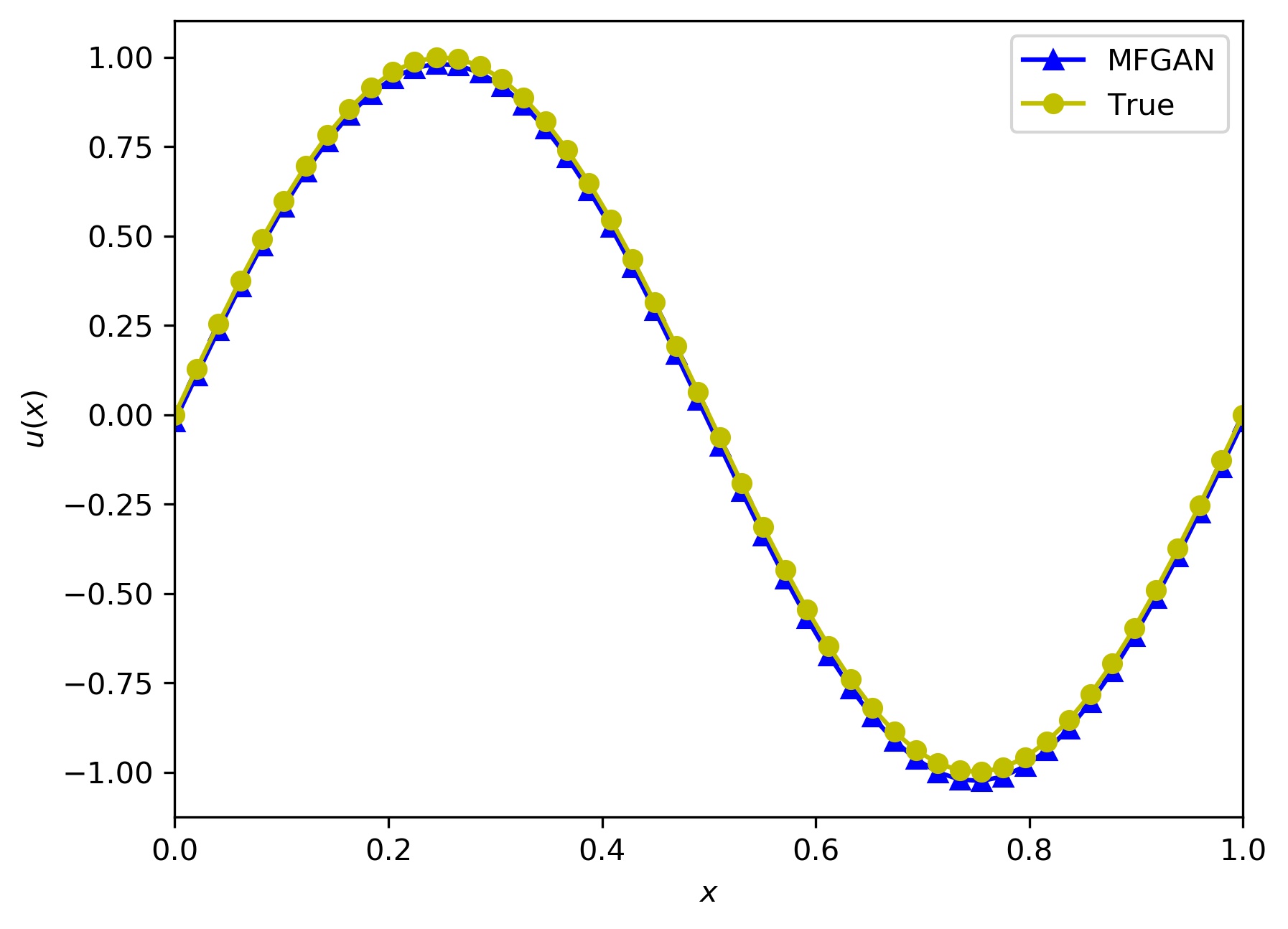}
 \subcaption{\centerline{\small Value function $u$.}
 \label{subfig:supp-1dim-u}}
 \end{minipage}
 \begin{minipage}[t]{0.4\columnwidth}
 \centering
 \includegraphics[width=\textwidth]{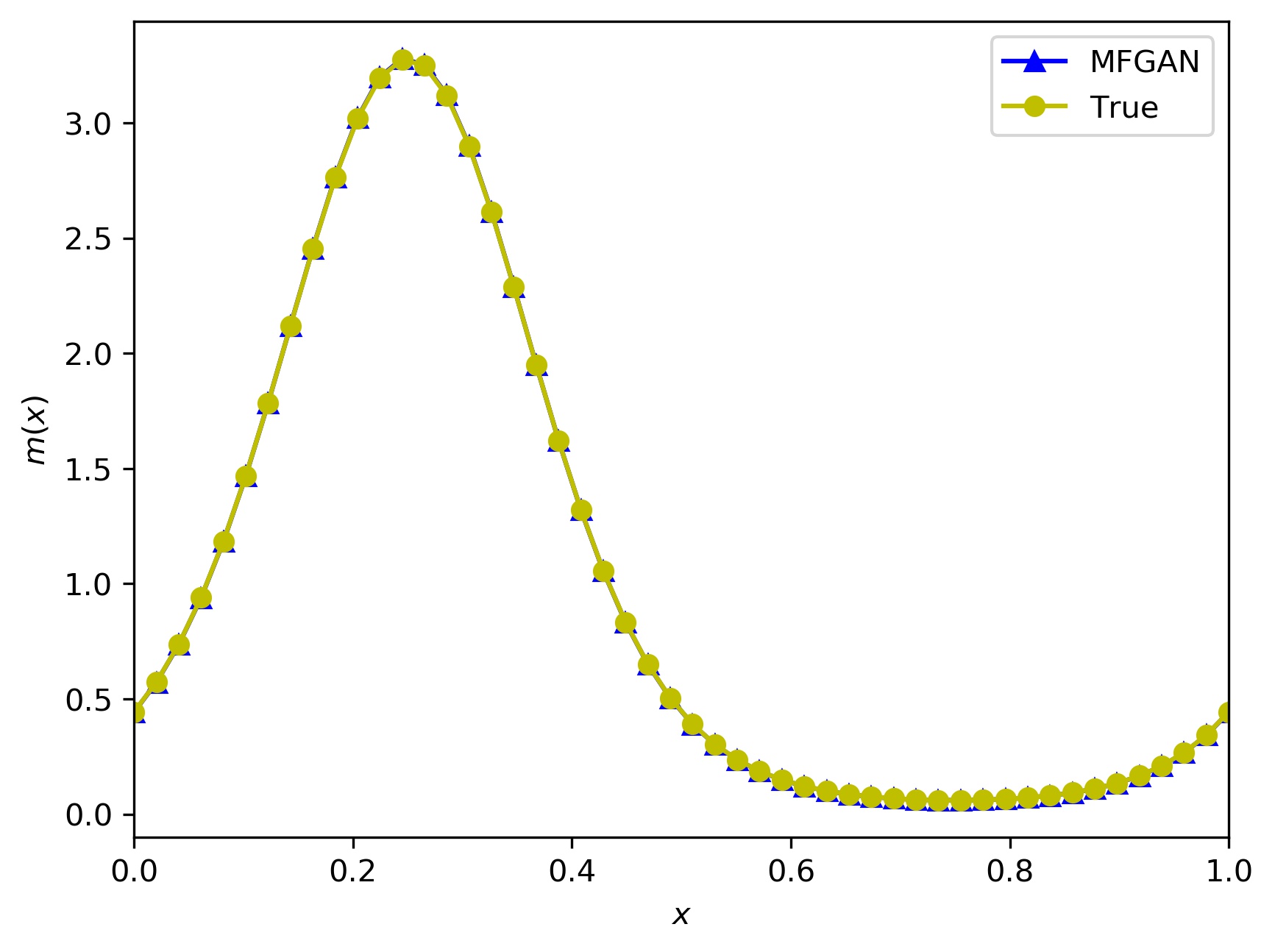}
 \subcaption{\centerline{\small Density function $m$.}
 \label{subfig:supp-1dim-m}}
 \end{minipage}\\
 \begin{minipage}[t]{0.4\columnwidth}
 \centering
 \includegraphics[width=\textwidth]{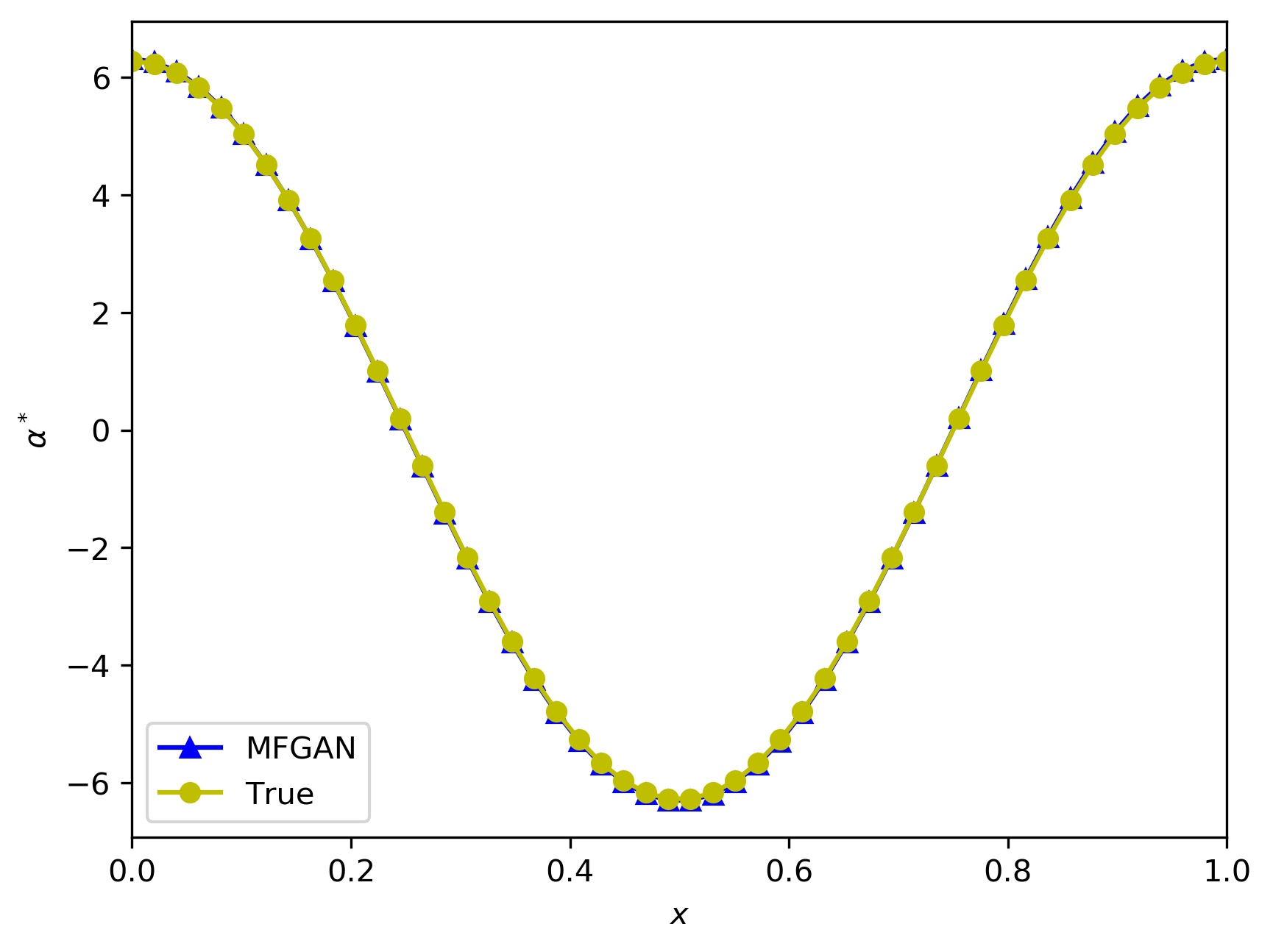}
 \subcaption{\centerline{\small Optimal control $\alpha^*$.}
 \label{subfig:supp-1dim-alpha}}
 \end{minipage}
 \caption{One-dimensional test case.}
 \label{fig:supp-1dim}
\end{figure}

\begin{figure}[!ht]
 \centering
 \begin{minipage}[t]{0.4\columnwidth}
 \centering
 \includegraphics[width=\textwidth]{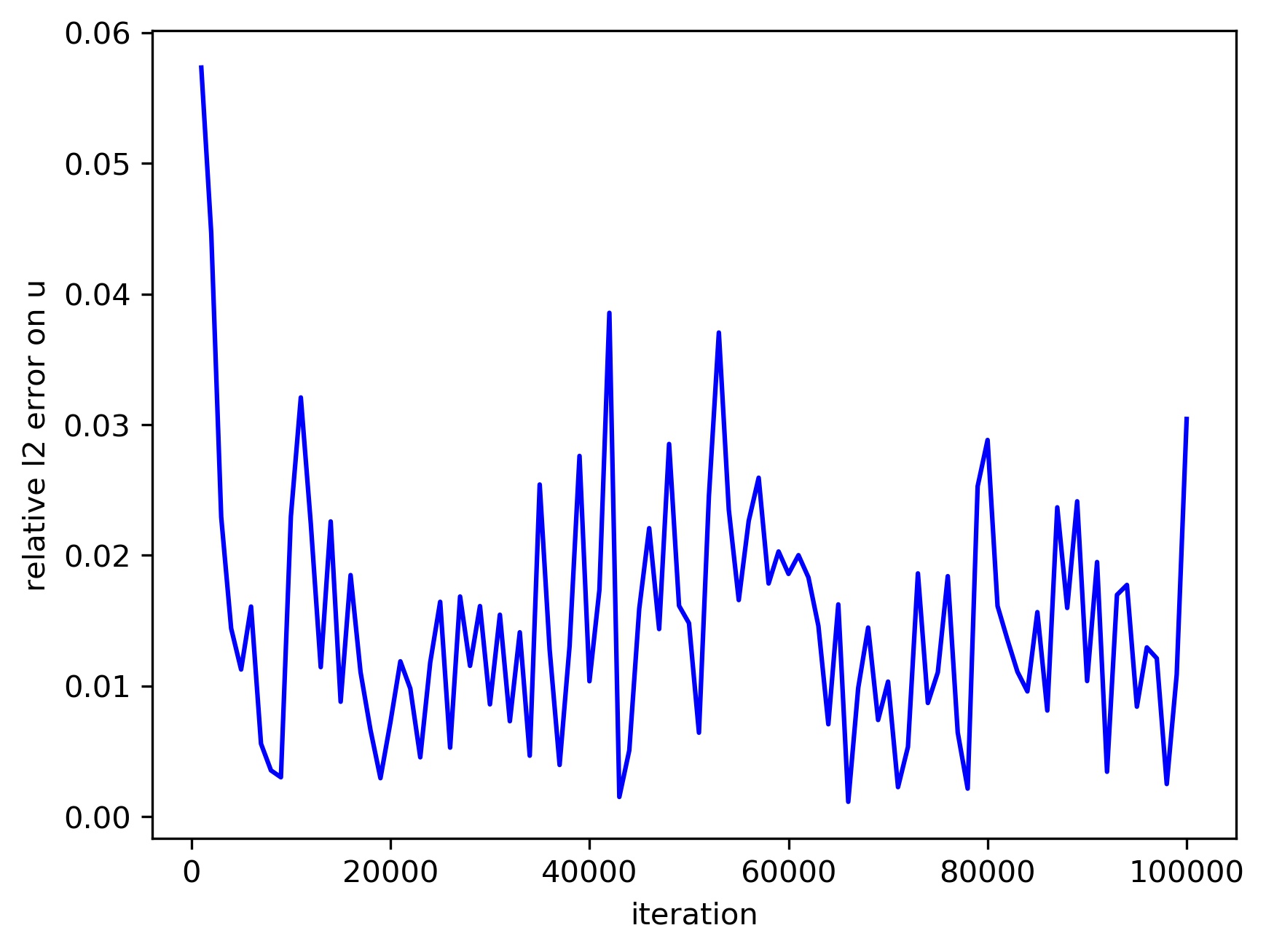}
 \subcaption{\centerline{\small Relative $l_2$ error of $u$.}
 \label{subfig:1dim-rel-err-u}}
 \end{minipage}
 \begin{minipage}[t]{0.4\columnwidth}
 \centering
 \includegraphics[width=\textwidth]{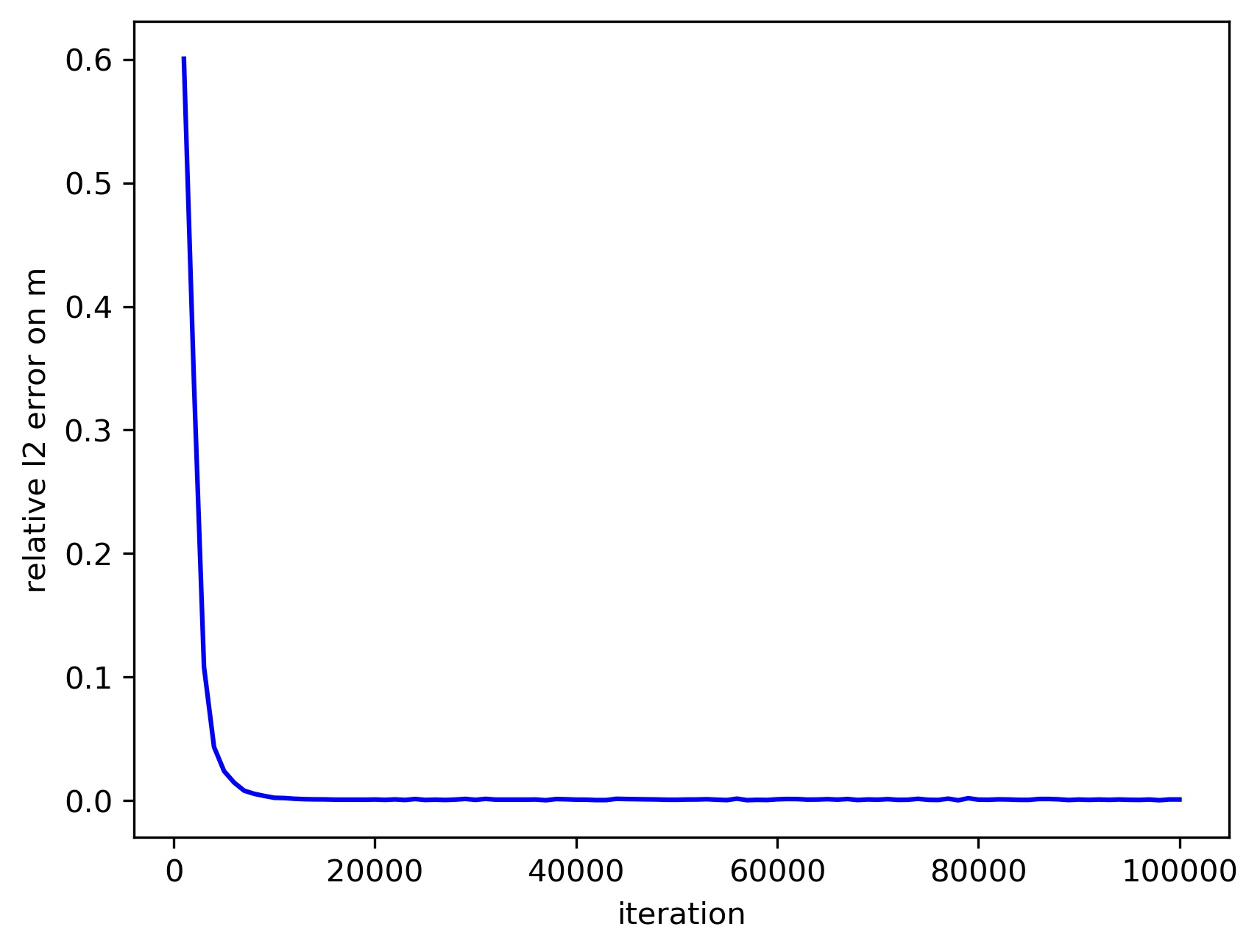}
 \subcaption{\centerline{\small Relative $l_2$ error of $m$.}
 \label{subfig:1dim-rel-err-m}}
 \end{minipage}\\
 \begin{minipage}[t]{0.4\columnwidth}
 \centering
 \includegraphics[width=\textwidth]{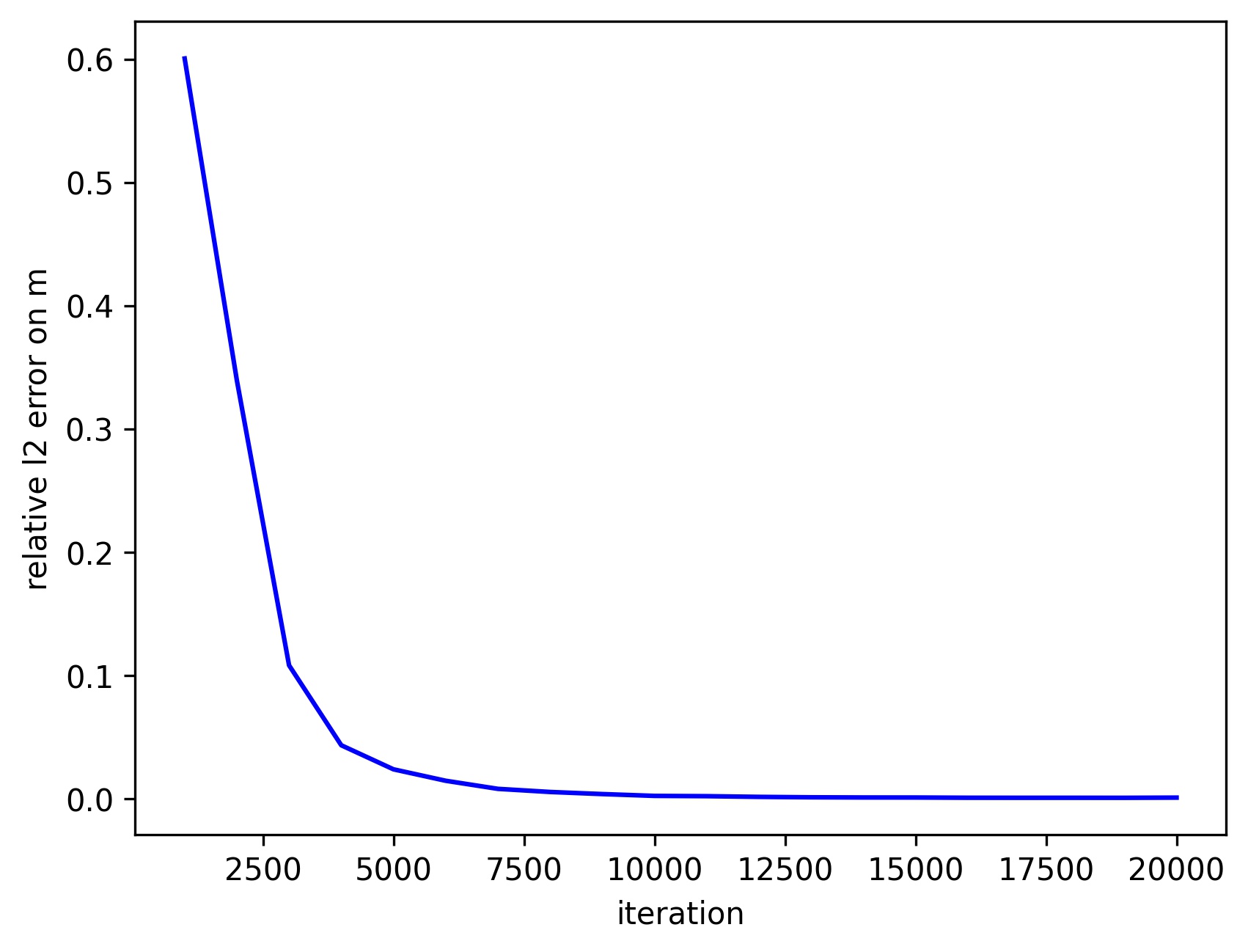}
 \subcaption{\centerline{\small Error of $m$ first 20k iterations.}
 \label{subfig:1dim-rel-err-m-dec}}
 \end{minipage}
 \begin{minipage}[t]{0.4\columnwidth}
 \centering
 \includegraphics[width=\textwidth]{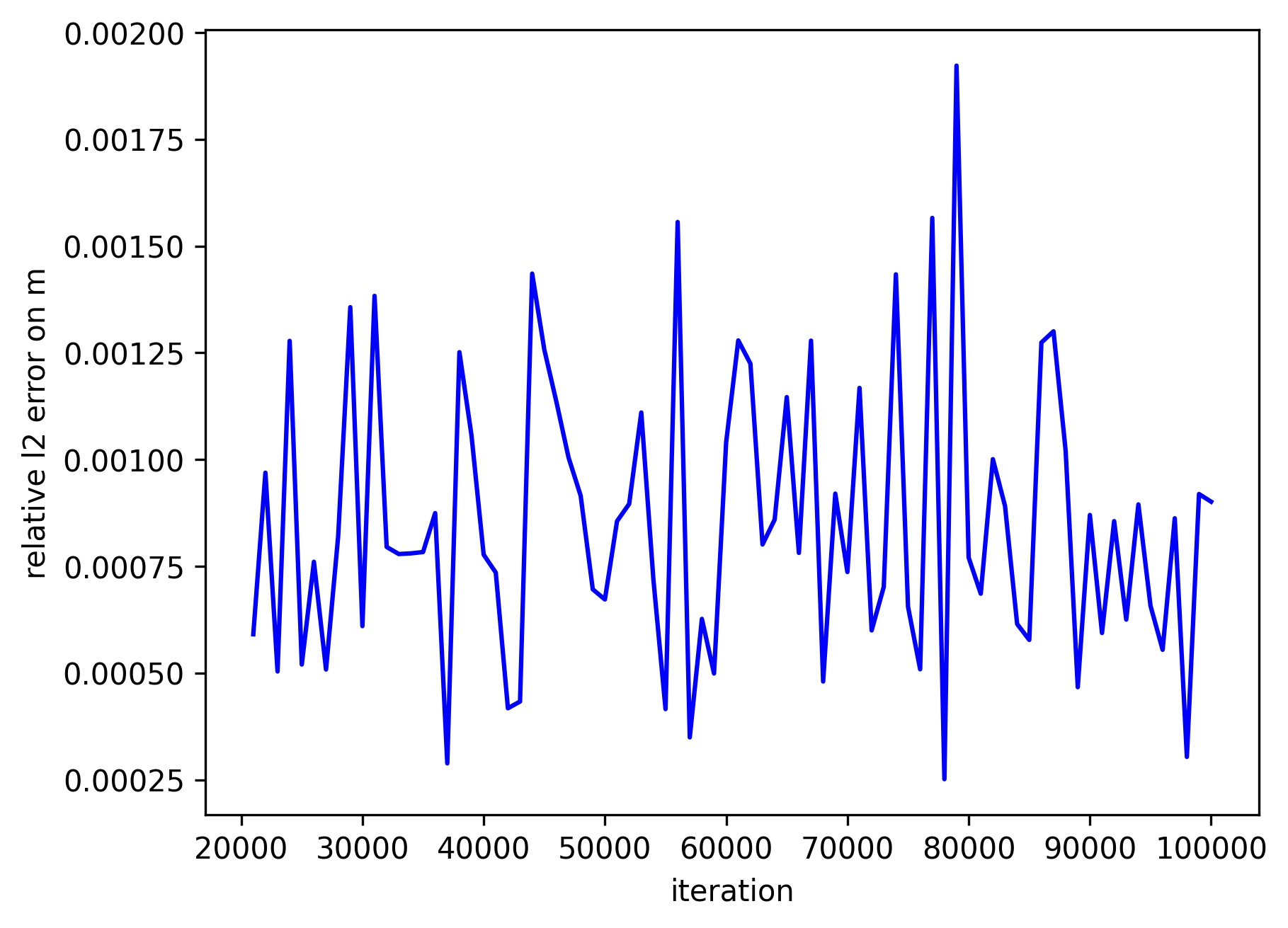}
 \subcaption{\centerline{\small Error of $m$ after 20k iterations.}
 \label{subfig:1dim-rel-err-m-flt}}
 \end{minipage}\\
 \begin{minipage}[t]{0.4\columnwidth}
 \centering
 \includegraphics[width=\textwidth]{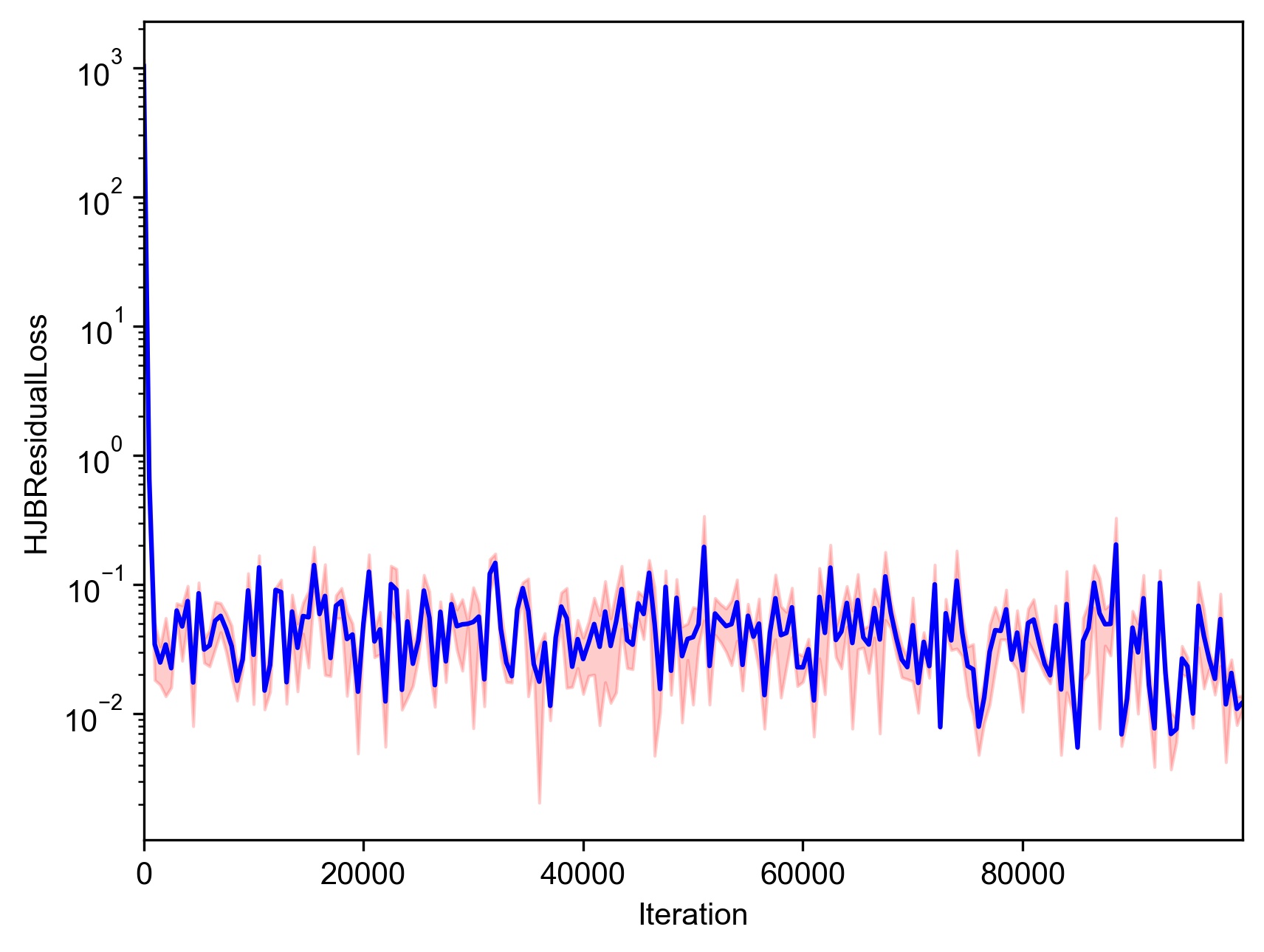}
 \subcaption{\centerline{\small (e) HJB residual loss.}
 \label{subfig:1dim-hjb}}
 \end{minipage}
 \begin{minipage}[t]{0.4\columnwidth}
 \centering
 \includegraphics[width=\textwidth]{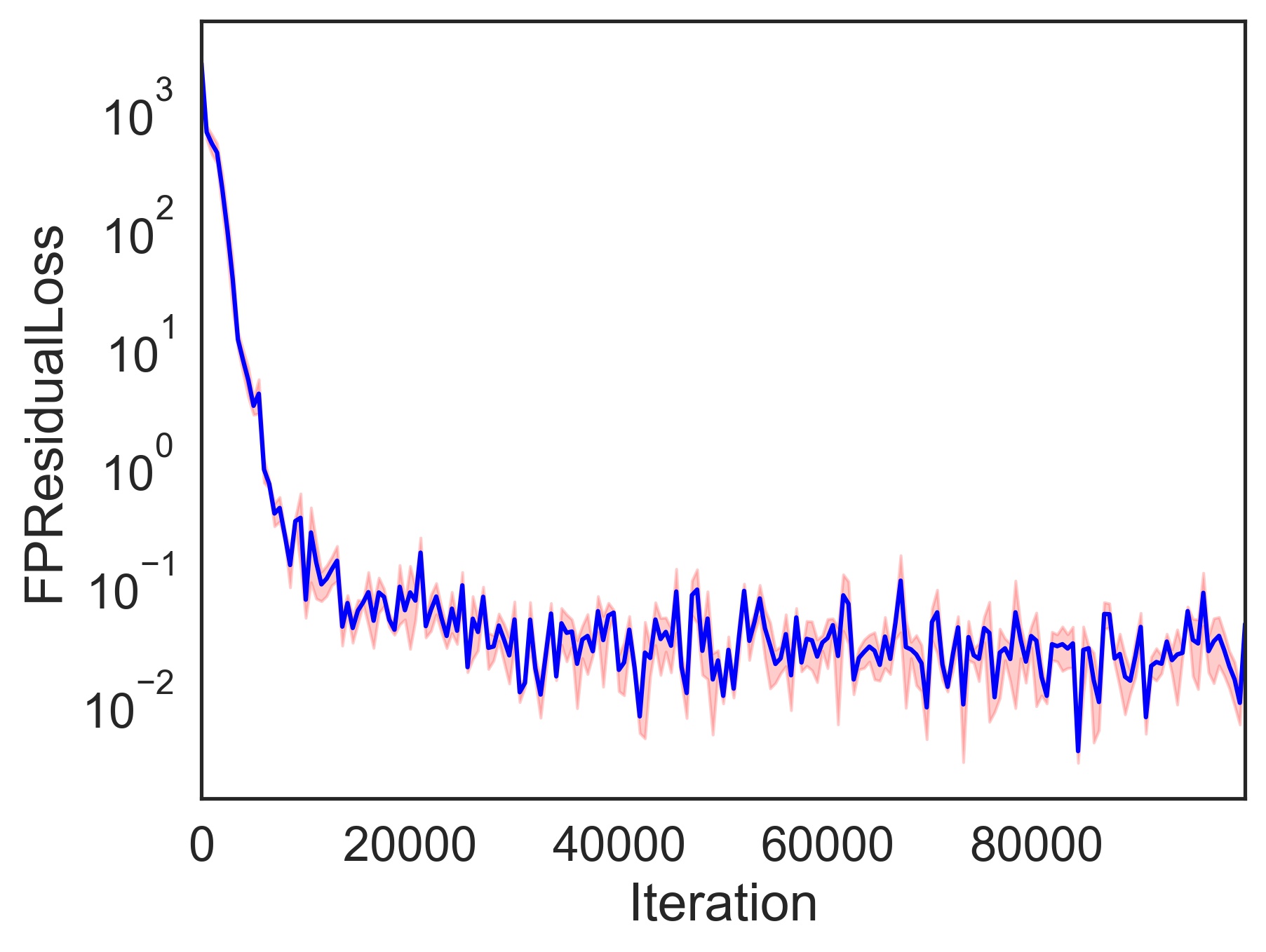}
 \subcaption{\centerline{\small (f) FP residual loss.}
 \label{subfig:1dim-fp}}
 \end{minipage}
 \caption{Losses and errors in the one-dimensional test case.}
 \label{fig:1dim}
\end{figure}

The Algorithm \ref{alg:mfgan-dyn} is first tested on a one-dimensional case, with its result highlighted in Figures \ref{fig:supp-1dim} and \ref{fig:1dim}. Figures \ref{subfig:supp-1dim-u} and \ref{subfig:supp-1dim-m} show the learnt functions of $u$ and $m$ against the true ones, respectively, and Figure \ref{subfig:supp-1dim-alpha} shows the optimal control, with the accuracy of the learnt functions versus the true ones. The plots of loss in Figures \ref{subfig:1dim-rel-err-u} and \ref{subfig:1dim-rel-err-m}, depict the evolution of relative $l_2$ error as the number of outer iterations grows to $K$. Within $10^5$ iterations, the relative $l_2$ error of $u$ oscillates around $3\times10^{-2}$, and the relative $l_2$ errors of $m$ decreases below $10^{-3}$. 
The evolution of the HJB and FP differential residual loss is shown in Figures \ref{subfig:1dim-hjb} and \ref{subfig:1dim-fp}, respectively. In theses figures, the solid line is the average loss among three experiments, with standard deviation captured by the shadow around the line. Both differential residuals first rapidly descend to the magnitude of $10^{-2}$ and then the descent slows down accompanied by oscillation.

\begin{figure}[!ht]
 \centering
 \begin{minipage}[t]{0.4\columnwidth}
 \centering
 \includegraphics[width=\textwidth]{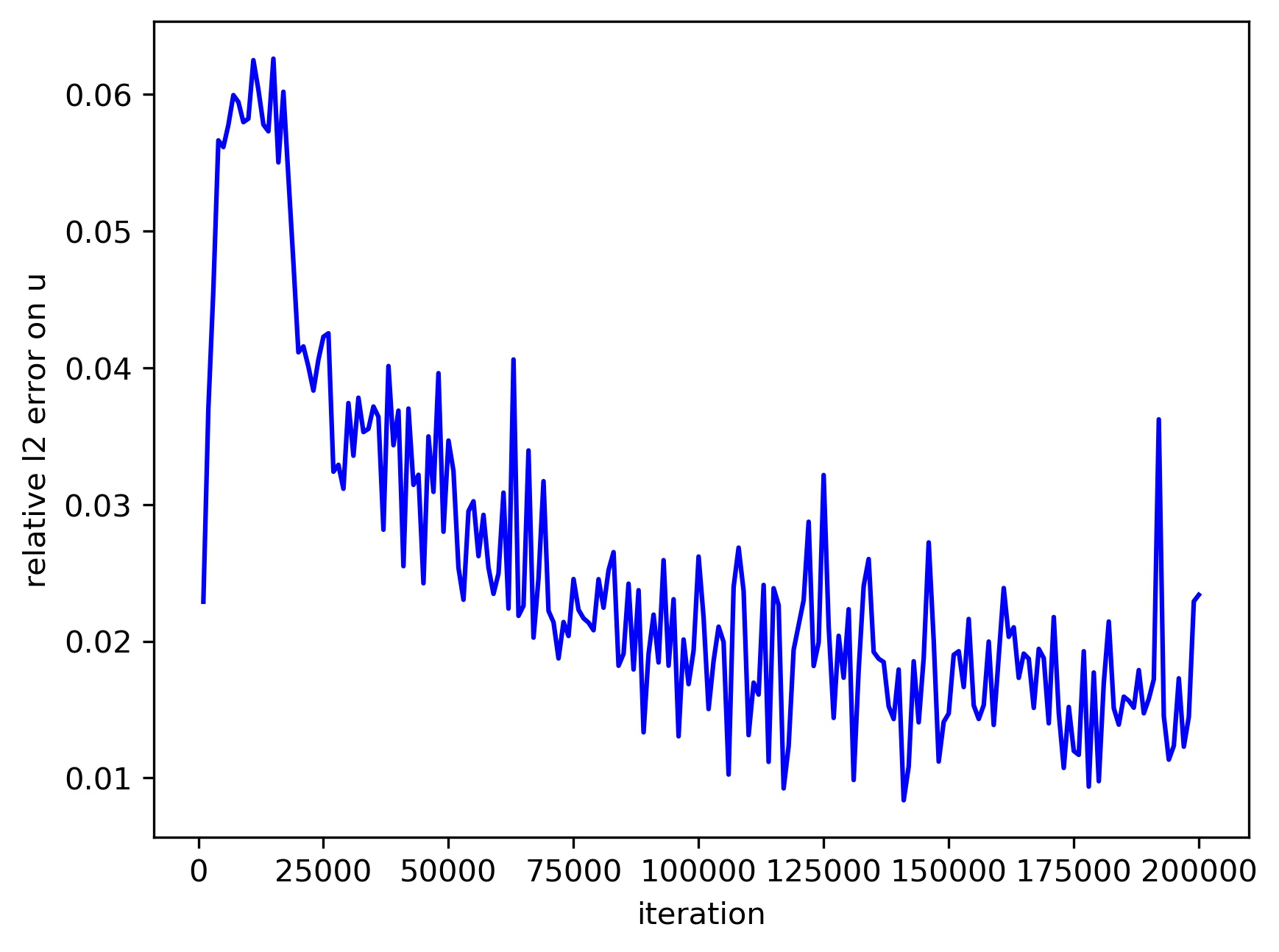}
 \subcaption{\centerline{\small Relative $l_2$ error of $u$.}
 \label{subfig:4dim-rel-err-u}}
 \end{minipage}
 \begin{minipage}[t]{0.4\columnwidth}
 \centering
 \includegraphics[width=\textwidth]{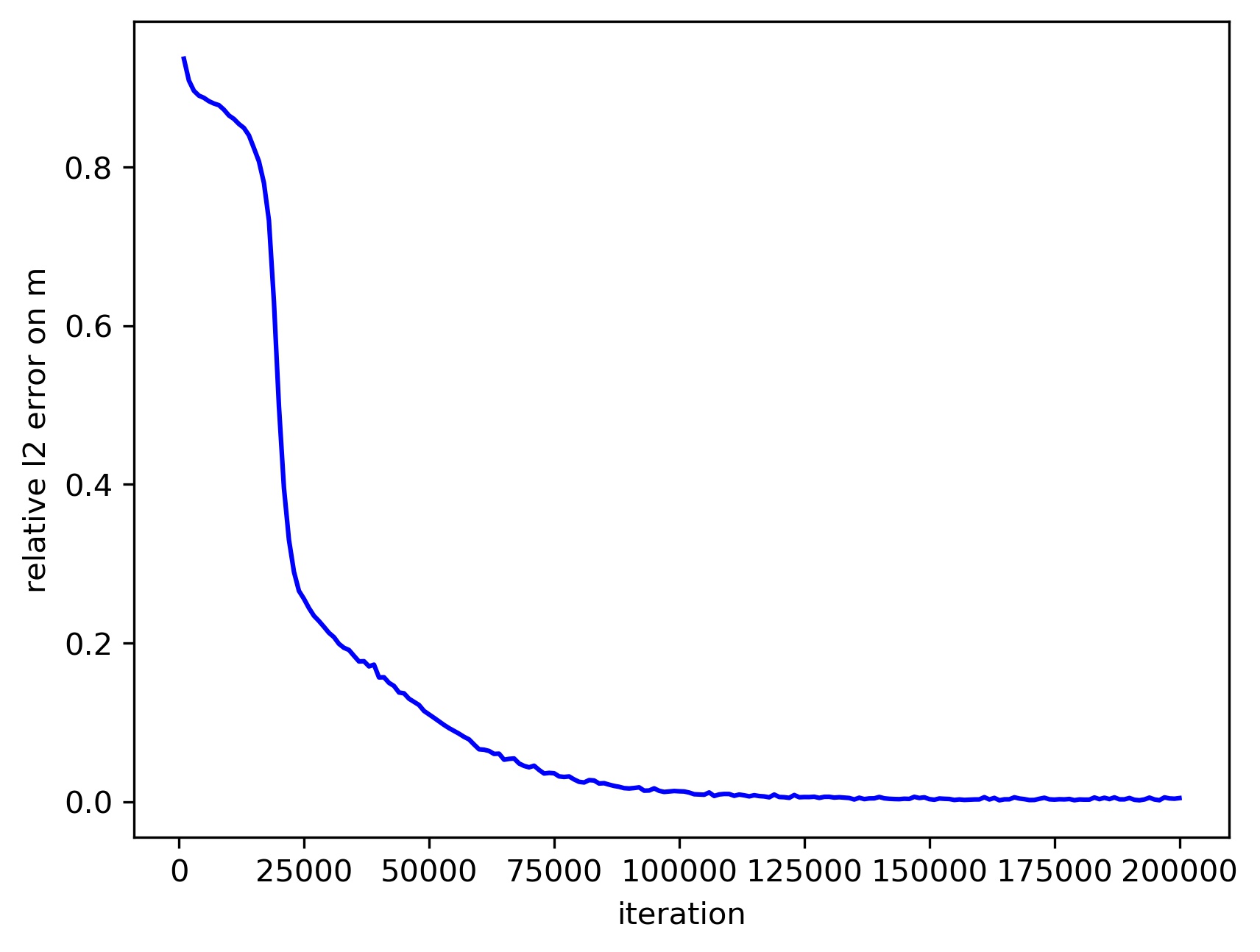}
 \subcaption{\centerline{\small Relative $l_2$ error of $m$.}
 \label{subfig:4dim-rel-err-m}}
 \end{minipage}\\
 \begin{minipage}[t]{0.4\columnwidth}
 \centering
 \includegraphics[width=\textwidth]{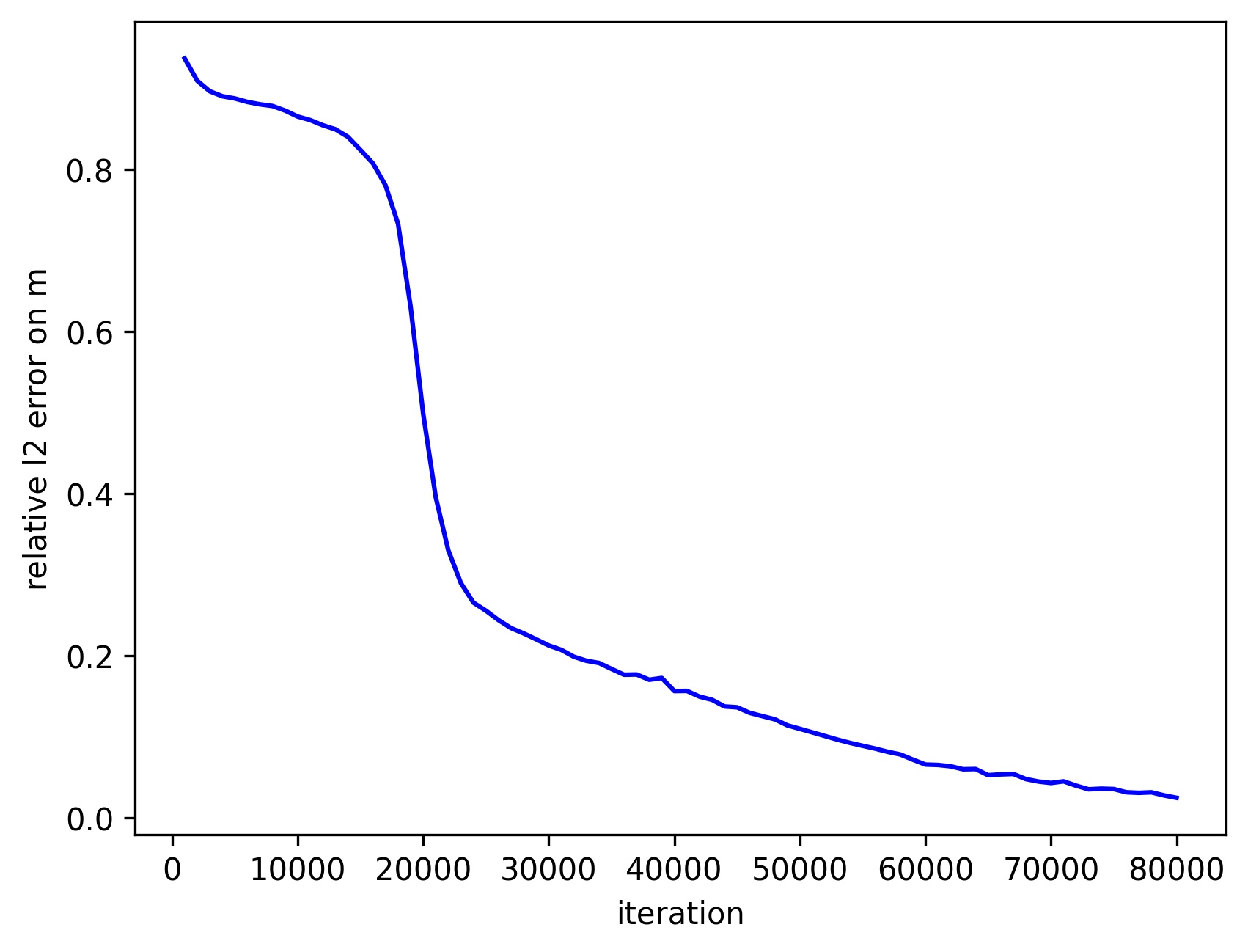}
 \subcaption{\centerline{\small Error of $m$ first 80k iterations.}
 \label{subfig:4dim-rel-err-m-dec}}
 \end{minipage}
 \begin{minipage}[t]{0.4\columnwidth}
 \centering
 \includegraphics[width=\textwidth]{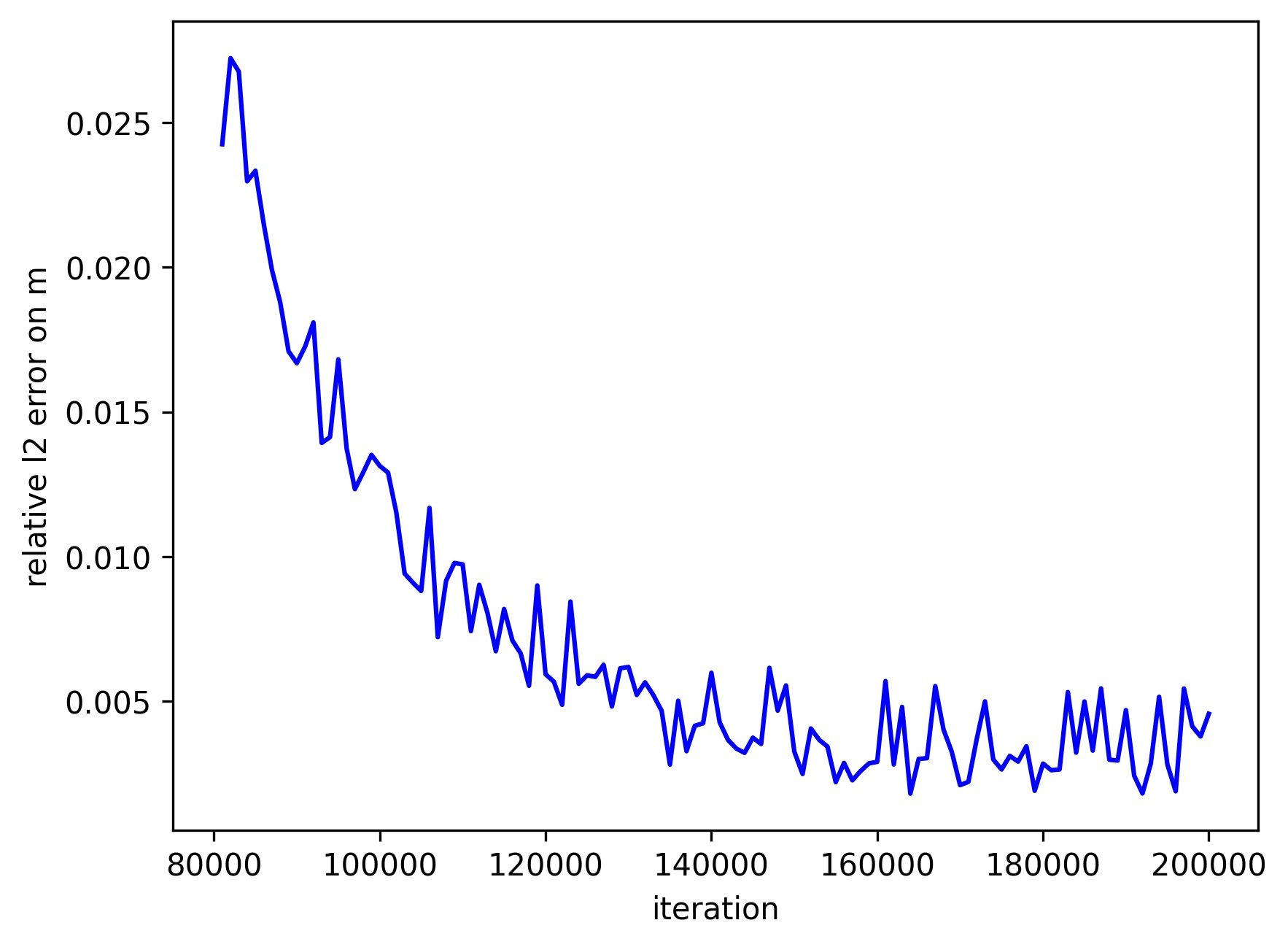}
 \subcaption{\centerline{\small Error of $m$ after 80k iterations.}
 \label{subfig:4dim-rel-err-m-flt}}
 \end{minipage}
 \caption{Input of dimension 4.}
 \label{fig:4dim}
\end{figure}

Algorithm \ref{alg:mfgan-dyn} is then applied to a four-dimensional case, with  result shown in Figure \ref{fig:4dim}.
Within $2\times 10^5$ iterations, the relative $l_2$ error of $u$ decreases below $2\times10^{-2}$ and that of $m$ decreases to $4\times10^{-3}$. 

Note the advantage of GANs training when compared with a similar experiment in Test Case 4 in \citep{CarmonaLauriere_DL} without the adversarial training for two neural networks: algorithms in \citep{CarmonaLauriere_DL} need significantly larger number of iterations: $10^6$ of iterations versus $2\times10^5$ for Algorithm \ref{alg:mfgan-dyn} to achieve the same level of accuracy.

A concurrent paper alongside with \citep{cao2020connecting} is the work of \citep{lin2020apac}. Using a primal-dual variational formulation associated with the coupled HJB-FP system as in \citep{Cirant2018}, MFGs are recast as GANs in a different way in \citep{lin2020apac}, where the density function is seen as the generator and the value function is seen as the discriminator. Based on this alternative interpretation, a GANs-based algorithm named APAC-Net is proposed. Through numerical experiments, this algorithm is shown to be able to solve certain classes of MFGs in dimension up to 100. 

\subsection{GANs in Mathematical Finance}
There are essentially two different frameworks in which GANs have been adopted in the mathematical finance literature. The first one is to reformulate a constrained control and optimization problem as a minimax problem so that the generator and discriminator networks  can be constructed for computational purpose. The second one is to draw the analogy between simulation of financial time series data and image generation such that various statistical and distributional properties can be exploited for performance evaluations. We will review several representative works for each category.

\subsubsection{Asset pricing and minimax problem.}
The work of \citep{Chen2019} is one of the earliest works to identify the minimax structure in a non-linear model for asset pricing. Its primary idea is to exploit the no-arbitrage condition and recast the constrained problem into the minimax framework of GANs. 
Their objective is to estimate the pricing kernel or stochastic discount factor (SDF) that summarizes the information of the cross-section of returns for different stocks. 

Specifically,  take the return of asset $i\in\{1,\dots,n\}$ at time $t+1$ as $R_{t+1,i}$ and the excess return as $R_{t+1,i}^e=R_{t+1,i} -R_{t,i}$. Let $M_{t+1}$ be the SDF satisfying the no-arbitrage condition,
\[
\mathbb E_t\left[M_{t+1}R_{t+1,i}^e\right]=0\Longleftrightarrow \mathbb E_t\left[R_{t+1,i}^e\right]=\left(-\frac{Cov_t(R_{t+1,i}^e, M_{t+1})}{Var_t(M_{t+1})}\right)\cdot\frac{Var_t(M_{t+1})}{\mathbb E_T\left[M_{t+1}\right]},
\]
where $\mathbb E_t$ stands for expectation conditional on some suitable information by time $t$. Then assume that
\[
M_{t+1}=1-\omega^TR_{t+1}^e=1-F_{t+1},\quad \beta_{t,i}=-\frac{Cov_t(R_{t+1,i}^e, M_{t+1})}{Var_t(M_{t+1})},
\]
where $\omega=(\omega_1,\dots,\omega_i,\dots,\omega_n)$ denotes the SDF weights vector which is also the weights vector of the conditional mean-variance efficient portfolio, and $\beta_{t,i}$ denotes the time-varying exposure to systematic risk for asset $i$.

In this one-factor model setup, the main quantities to be estimated are the two vectors $\omega$ and $\beta_t=(\beta_{t,1},\dots,\beta_{t,i},\dots,\beta_{t,n})$.
To handle the no-arbitrage constraint, they utilize the unconditional moment conditions: given any $\sigma$-algebra generated by some random variable $Z$, $\mathcal F=\sigma(Z)$, 
\[Y=\mathbb E[X|\mathcal F]\Longrightarrow\mathbb E\left[Xf(Z)\right]=\mathbb E[Yf(Z)],
\] 
for any measurable function $f$. In particular, let the choice of information be $\sigma(I_t, I_{t,i})$, where $I_t$ represents the macroeconomic conditions at time $t$ whereas $I_{t,i}$ denotes information at time $t$ for the specific stock $i$, then 
\[\omega_i=\omega_{t,i}=\omega(I_t,I_{t,i}),\quad \beta_{t,i}=\beta(I_t,I_{t,i}).\]
Consequently, the no-arbitrage condition implies
\[\mathbb E\left[M_{t+1}R_{t+1}^eg(I_t,I_{t,i})\right]=0,\]
for any measurable function $g$; if $\hat\omega$ and $\hat\beta$ correspond to the correct SDF $\hat M$, this is equivalent to
\[\max_{g}\frac{1}{N}\sum_{j=1}^N\left\|\mathbb E\left[\hat M_{t+1}R_{t+1,j}^eg(I_t,I_{t,j})\right]\right\|^2=0.\]
Now, estimating SDF that satisfies the no-arbitrage condition is transformed into the minimax game
\[\min_\omega\max_g\frac{1}{N}\sum_{j=1}^N\left\|\mathbb E\left[\hat M_{t+1}R_{t+1,j}^eg(I_t,I_{t,j})\right]\right\|^2,\]
a natural GANs structure. 

This proposed GANs model is then compared with an alternative model with the no-arbitrage condition relaxed to a first moment condition given by the one-factor model, $\mathbb E[R_{t+1,j}^e]\propto \mathbb E[F_{t+1}]$. It is further compared with a second alternative model with both $\omega$ and $g$ assumed to be linear. The GANs model is shown to outperform uniformly in terms of Sharp ratio, explained variation, and cross-sectional mean $R^2$.

\subsubsection{GANs as financial time series data simulators}
Another application of GANs is to generate financial time series data for both equity and derivatives.

In \citep{Wiese2019}, the main objective is to build a simulator for equity option markets. Instead of dealing with option price directly which is subject to the no-arbitrage constraint, they work with an equivalent and less constrained form called discrete local volatility (DLV). 

In this formulation, the time-varying DLV $\sigma_t$ is seen as a function of strike $K$ and maturity $M$. The generator takes the state variable $S_t=f(\sigma_t,\dots, \sigma_0)$ as well as some random noise $Z_{t+1}$ as inputs and set $X_{t+1} = \log \sigma_{t+1}=g(Z_{t+1}, S_t)$.
The discriminator tries to distinguish the true $(X_{t+1},S_t)$ and the generated $(\tilde X_{t+1}, \tilde S_t)$. Other calibration techniques such as PCA are also incorporated.

This formulation is compared among different neural network based simulators. The performance evaluation is based on four types of criteria: the distributional metric which is the distance between the empirical probability distribution functions of the generated and historical data, the distributional scores given by skewness and kurtosis scores, the dependency score through the autocorrelation function score for the log-return process and finally the cross-correlation scores for the log-DLV and the DLV log returns. Their numerical results show that the GANs model outperforms the other benchmark models such as vector autoregressive models, TCN models, and quasi maximum likelihood estimation.

In a closely related work, \citep{Wiese2020} proposes a special GANs model called the Quant GAN. The main characteristic of Quant GANs is taking temporal convolutional networks (TCNs) as the generator. By choosing appropriate kernel size $K$ and dilation factor $D$, TCNs can carry long-time dependency and avoid abnormal behavior of gradients over time. They show that with Lipschitz constraint on the choices of activation functions and weights, the generated process has as many number of moments as the input latent variable. Finally, they use the inverse Lambert $W$ transform for the real asset log-return processes to copy with the heavy-tail property in the GANs training. In the Lambert $W$ transform, a random variable $X$ with mean $\mu$, variance $\sigma^2$ and cumulative distribution function $F_X$ is transformed into
\[Y=\frac{X-\mu}{\sigma}\exp\left(\frac{\delta(X-\mu)^2}{2\sigma^2}\right)+\mu,\]
with a proper choice of nonnegative parameter $\delta$ so that $Y$ has heavier tail than $X$ if $\delta>0$. The inverse Lambert $W$ transform is its inverse process.

They propose two different approaches of utilizing TCNs: one is to use the pure TCNs to directly generate time series, the other is to use TCNs to generate drift and volatility process and add another network to represent the noise. They test the simple GARCH model for comparison, with the evaluation of the models based on distributional metrics and dependence scores. In particular, the former include Wasserstein distance and DY metric, i.e., a measurement of the distance between the estimated likelihoods from real and generated data, and the latter include ACF score and the leverage effect score. Their results show that the GANs model with pure TCNs perform the best for the majority of the tests, and that both GANs models dominate the GARCH model.

Other related works include \citep{Takahashi2019} and \citep{Zhang2019}. The GANs model in \citep{Takahashi2019} captures statistical properties exhibited in real financial data, such as linear unpredictability, the heavy-tailed price return distribution, volatility clustering, leverage effects, the coarse-fine volatility correlation, and the gain/loss asymmetry. The GANs model in \citep{Zhang2019} is used to predict stock prices from historical stock data, where long-short-term-memory is adopted as the generator and multi-layer perceptron as the discriminator. In particular, the generator acts as a function characterizing the unknown and possibly complex relation between stock price in the future and historical data.

There are other extensions of GANs models. For instance, in \citep{cao2020conditional}, conditional GANs are constructed to simulate quantities that have traditionally been of interest in financial industry. This GANs model enables dynamic data updating for stress tests. Embracing the general idea of adversarial training in GANs, \citep{cuchiero2020generative} proposes a generative adversarial approach for (robust) calibration of local stochastic volatility models; the generation of volatility surfaces follows neural SDEs, where a special deep-hedging-based variance reduction technique is applied and the adversarial training idea is embedded in evaluating the simulated volatility surfaces: the loss function may come from a family of candidate loss functions to ensure robustness. Recently, a GANs model called COT-GAN is proposed in \citep{xu2020cot} based on causal optimal transport theory. In this work, the temporal causality condition naturally leads to an adversarial framework for GANs and a mixed Sinkhorn distance is proposed to calculate the optimal transport cost with reduced bias. This new framework could be used for generating sequential data including financial time series.

\section{Conclusion and Discussion}\label{sec: ccl}
This notes covers three major aspects of GANs,  essentials of GANs in the optimization and game framework, GANs training via stochastic analysis, and recent applications of GANs in mathematical finance. 

Despite its vast popularity and power in data and image generation,
GANs face many challenges in implementation and training and remain
largely undeveloped in theory. For instance, the well-posedness of GANs as a minimax game has not been fully understood until \cite{guo2020optimal} in which the 
convexity issue is analyzed in details. The connection between 
mean-field games and GANs via the minimax structure presents GANs' potential  computing power for high dimensional control and optimization problems with variational structures. 
The next natural test field is forward-backward-stochastic-differential equations, where there is a natural variational structure to retrofit for the minimax game of GANs . Beyond computational power, more explorations are needed to see if convergence and computation complexity results can be obtained, especially given the SDE approximation of GANs training.
A small step towards this direction is \citep{guo2020optimal}, which formulates  simple stochastic control problems for learning rate and batch size analysis and shows their impact on error and variance reduction. One also wonders if the empirical success of GANs in data generation can be replicated in the general area of simulation and if robust theoretical analysis can be established. 

\bibliographystyle{apalike}
\bibliography{yourchapter} 

\begin{thebibliography}{}

\bibitem[Arjovsky and Bottou, 2017]{Arjovsky2017a}
Arjovsky, M. and Bottou, L. (2017).
\newblock {Towards principled methods for training generative adversarial
  networks}.
\newblock In {\em International Conference on Learning Representations},
  Toulon.

\bibitem[Arjovsky et~al., 2017]{Arjovsky2017}
Arjovsky, M., Chintala, S., and Bottou, L. (2017).
\newblock {Wasserstein generative adversarial networks}.
\newblock In {\em International Conference on Machine Learning}, pages
  214--223.

\bibitem[Barnett, 2018]{Barnett2018}
Barnett, S.~A. (2018).
\newblock {Convergence problems with generative adversarial networks (GANs)}.
\newblock {\em arXiv preprint arXiv:1806.11382}.

\bibitem[Berard et~al., 2020]{Berard2020}
Berard, H., Gidel, G., Almahairi, A., Vincent, P., and Lacoste-Julien, S.
  (2020).
\newblock {A closer look at the optimization landscape of generative
  adversarial networks}.
\newblock In {\em International Conference on Learning Representations}.

\bibitem[Cao and Guo, 2020]{cao2020approximation}
Cao, H. and Guo, X. (2020).
\newblock {Approximation and convergence of GANs training: an SDE approach}.
\newblock {\em arXiv preprint arXiv:2006.02047}.

\bibitem[Cao et~al., 2020a]{cao2020connecting}
Cao, H., Guo, X., and Lauri{\`e}re, M. (2020a).
\newblock {Connecting GANs, MFGs and OT}.
\newblock {\em arXiv preprint arXiv:2002.04112}.

\bibitem[Cao et~al., 2020b]{cao2020conditional}
Cao, H., Guo, X., and Lehalle, C.-A. (2020b).
\newblock {Conditional GANs and stress testing}.
\newblock {\em Preprint}.

\bibitem[Carmona and Lauri{\`e}re, 2019]{CarmonaLauriere_DL}
Carmona, R. and Lauri{\`e}re, M. (2019).
\newblock Convergence analysis of machine learning algorithms for the numerical
  solution of mean field control and games: {II} - the finite horizon case.
\newblock Preprint.

\bibitem[Chen et~al., 2019]{Chen2019}
Chen, L., Pelger, M., and Zhu, J. (2019).
\newblock {Deep learning in asset pricing}.
\newblock {\em Available at SSRN 3350138}.

\bibitem[Cirant and Nurbekyan, 2018]{Cirant2018}
Cirant, M. and Nurbekyan, L. (2018).
\newblock {The variational structure and time-periodic solutions for mean-field
  games systems}.
\newblock {\em arXiv preprint arXiv:1804.08943}.

\bibitem[Conforti et~al., 2020]{Conforti2020}
Conforti, G., Kazeykina, A., and Ren, Z. (2020).
\newblock {Game on random environment, mean-field Langevin system and neural
  networks}.
\newblock {\em arXiv preprint arXiv:2004.02457}.

\bibitem[Cuchiero et~al., 2020]{cuchiero2020generative}
Cuchiero, C., Khosrawi, W., and Teichmann, J. (2020).
\newblock A generative adversarial network approach to calibration of local
  stochastic volatility models.
\newblock {\em Risks}, 8(4):101.

\bibitem[Denton et~al., 2015]{denton2015deep}
Denton, E.~L., Chintala, S., Szlam, A., and Fergus, R. (2015).
\newblock Deep generative image models using a {Laplacian} pyramid of
  adversarial networks.
\newblock In {\em Advances in Neural Information Processing Systems}, pages
  1486--1494.

\bibitem[Domingo-Enrich et~al., 2020]{Domingo-Enrich2020}
Domingo-Enrich, C., Jelassi, S., Mensch, A., Rotskoff, G.~M., and Bruna, J.
  (2020).
\newblock {A mean-field analysis of two-player zero-sum games}.
\newblock {\em arXiv preprint arXiv:2002.06277}.

\bibitem[Goodfellow et~al., 2014]{Goodfellow2014}
Goodfellow, I.~J., Pouget-Abadie, J., Mirza, M., Xu, B., Warde-Farley, D.,
  Ozair, S., Courville, A., and Bengio, Y. (2014).
\newblock {Generative adversarial nets}.
\newblock In {\em Advances in Neural Information Processing Systems}, pages
  2672--2680.

\bibitem[Guo et~al., 2017]{guo2017relaxed}
Guo, X., Hong, J., Lin, T., and Yang, N. (2017).
\newblock Relaxed {Wasserstein} with applications to {GANs}.
\newblock {\em arXiv preprint arXiv:1705.07164}.

\bibitem[Guo and Mounjid, 2020]{guo2020optimal}
Guo, X. and Mounjid, O. (2020).
\newblock {Optimal learning rate for GANs via SDEs}.
\newblock {\em Preprint}.

\bibitem[Heusel et~al., 2017]{Heusel2017}
Heusel, M., Ramsauer, H., Unterthiner, T., Nessler, B., and Hochreiter, S.
  (2017).
\newblock {GANs trained by a two time-scale update rule converge to a local
  Nash equilibrium}.
\newblock In {\em Advances in Neural Information Processing Systems}, pages
  6626--6637.

\bibitem[Kulharia et~al., 2017]{ghosh2016contextual}
Kulharia, V., Ghosh, A., Mukerjee, A., Namboodiri, V., and Bansal, M. (2017).
\newblock Contextual {RNN-GANs} for abstract reasoning diagram generation.
\newblock In {\em Proceedings of the Thirty-First AAAI Conference on Artificial
  Intelligence}, pages 1382--1388.

\bibitem[Ledig et~al., 2017]{ledig2016others}
Ledig, C., Theis, L., Husz{\'a}r, F., Caballero, J., Cunningham, A., Acosta,
  A., Aitken, A., Tejani, A., Totz, J., Wang, Z., et~al. (2017).
\newblock Photo-realistic single image super-resolution using a generative
  adversarial network.
\newblock In {\em Proceedings of the IEEE Conference on Computer Vision and
  Pattern Recognition}, pages 4681--4690.

\bibitem[Lei et~al., 2019]{lei2019geometric}
Lei, N., Su, K., Cui, L., Yau, S.-T., and Gu, X.~D. (2019).
\newblock A geometric view of optimal transportation and generative model.
\newblock {\em Computer Aided Geometric Design}, 68:1--21.

\bibitem[Li et~al., 2019]{Li2019}
Li, Q., Tai, C., and E, W. (2019).
\newblock {Stochastic modified equations and dynamics of stochastic gradient
  algorithms I: mathematical foundations}.
\newblock {\em Journal of Machine Learning Research}, 20(40):1--47.

\bibitem[Lin et~al., 2020]{lin2020apac}
Lin, A.~T., Fung, S.~W., Li, W., Nurbekyan, L., and Osher, S.~J. (2020).
\newblock {APAC-Net: Alternating the population and agent control via two
  neural networks to solve high-dimensional stochastic mean field games}.
\newblock {\em arXiv preprint arXiv:2002.10113}.

\bibitem[Liu and Theodorou, 2019]{Liu2019b}
Liu, G.-H. and Theodorou, E.~A. (2019).
\newblock {Deep learning theory review: An optimal control and dynamical
  systems perspective}.
\newblock {\em arXiv preprint arXiv:1908.10920}.

\bibitem[Luc et~al., 2016]{luc2016semantic}
Luc, P., Couprie, C., Chintala, S., and Verbeek, J. (2016).
\newblock Semantic segmentation using adversarial networks.
\newblock {\em arXiv preprint arXiv:1611.08408}.

\bibitem[Mescheder et~al., 2018]{Mescheder2018}
Mescheder, L., Geiger, A., and Nowozin, S. (2018).
\newblock {Which training methods for GANs do actually converge?}
\newblock In {\em International Conference on Machine Learning}, pages
  3481----3490.

\bibitem[Nock et~al., 2017]{nock2017f}
Nock, R., Cranko, Z., Menon, A.~K., Qu, L., and Williamson, R.~C. (2017).
\newblock f-{GAN}s in an information geometric nutshell.
\newblock In {\em Advances in Neural Information Processing Systems}, pages
  456--464.

\bibitem[Nowozin et~al., 2016]{nowozin2016f}
Nowozin, S., Cseke, B., and Tomioka, R. (2016).
\newblock {f-GAN: training generative neural samplers using variational
  divergence minimization}.
\newblock In {\em Proceedings of the 30th International Conference on Neural
  Information Processing Systems}, pages 271--279.

\bibitem[Radford et~al., 2015]{radford2015unsupervised}
Radford, A., Metz, L., and Chintala, S. (2015).
\newblock Unsupervised representation learning with deep convolutional
  generative adversarial networks.
\newblock {\em arXiv preprint arXiv:1511.06434}.

\bibitem[Reed et~al., 2016]{reed2016generative}
Reed, S., Akata, Z., Yan, X., Logeswaran, L., Schiele, B., and Lee, H. (2016).
\newblock Generative adversarial text to image synthesis.
\newblock In {\em 33rd International Conference on Machine Learning}, pages
  1060--1069.

\bibitem[Salimans et~al., 2016]{Salimans2016}
Salimans, T., Goodfellow, I., Zaremba, W., Cheung, V., Radford, A., and Chen,
  X. (2016).
\newblock {Improved techniques for training GANs}.
\newblock In {\em Advances in Neural Information Processing Systems}, pages
  2234--2242.

\bibitem[Salimans et~al., 2018]{salimans2018improving}
Salimans, T., Zhang, H., Radford, A., and Metaxas, D. (2018).
\newblock Improving {GAN}s using optimal transport.
\newblock In {\em International Conference on Learning Representations}.

\bibitem[Sanjabi et~al., 2018]{sanjabi2018convergence}
Sanjabi, M., Ba, J., Razaviyayn, M., and Lee, J.~D. (2018).
\newblock On the convergence and robustness of training {GAN}s with regularized
  optimal transport.
\newblock In {\em Advances in Neural Information Processing Systems}, pages
  7091--7101.

\bibitem[Sion, 1958]{sion1958general}
Sion, M. (1958).
\newblock On general minimax theorems.
\newblock {\em Pacific Journal of Mathematics}, 8(1):171--176.

\bibitem[Srivastava et~al., 2019]{srivastava2019bregmn}
Srivastava, A., Greenewald, K., and Mirzazadeh, F. (2019).
\newblock {BreGMN}: scaled-{Bregman} generative modeling networks.
\newblock {\em arXiv preprint arXiv:1906.00313}.

\bibitem[Takahashi et~al., 2019]{Takahashi2019}
Takahashi, S., Chen, Y., and Tanaka-Ishii, K. (2019).
\newblock {Modeling financial time-series with generative adversarial
  networks}.
\newblock {\em Physica A: Statistical Mechanics and its Applications},
  527:121261.

\bibitem[Von~Neumann, 1959]{von1959theory}
Von~Neumann, J. (1959).
\newblock On the theory of games of strategy.
\newblock {\em Contributions to the Theory of Games}, 4:13--42.

\bibitem[Vondrick et~al., 2016]{vondrick2016generating}
Vondrick, C., Pirsiavash, H., and Torralba, A. (2016).
\newblock Generating videos with scene dynamics.
\newblock In {\em Advances in Neural Information Processing Systems}, pages
  613--621.

\bibitem[Wiatrak et~al., 2019]{Wiatrak2019}
Wiatrak, M., Albrecht, S.~V., and Nystrom, A. (2019).
\newblock {Stabilizing generative adversarial networks: a survey}.
\newblock {\em arXiv preprint arXiv:1910.00927}.

\bibitem[Wiese et~al., 2019]{Wiese2019}
Wiese, M., Bai, L., Wood, B., Morgan, J.~P., and Buehler, H. (2019).
\newblock {Deep hedging: learning to simulate equity option markets}.
\newblock {\em arXiv preprint arXiv:1911.01700}.

\bibitem[Wiese et~al., 2020]{Wiese2020}
Wiese, M., Knobloch, R., Korn, R., and Kretschmer, P. (2020).
\newblock {Quant GANs: deep generation of financial time series}.
\newblock {\em Quantitative Finance}, pages 1--22.

\bibitem[Xu et~al., 2020]{xu2020cot}
Xu, T., Wenliang, L.~K., Munn, M., and Acciaio, B. (2020).
\newblock {COT-GAN: Generating sequential data via causal optimal transport}.
\newblock {\em arXiv preprint arXiv:2006.08571}.

\bibitem[Yaida, 2019]{Yaida2019}
Yaida, S. (2019).
\newblock {Fluctuation-dissipation relations for stochastic gradient descent}.
\newblock In {\em International Conference on Learning Representations}.

\bibitem[Yang et~al., 2020]{Yang2018}
Yang, L., Zhang, D., and Karniadakis, G.~E. (2020).
\newblock Physics-informed generative adversarial networks for stochastic
  differential equations.
\newblock {\em SIAM Journal on Scientific Computing}, 42(1):A292--A317.

\bibitem[Yang and Perdikaris, 2019]{Yang2018a}
Yang, Y. and Perdikaris, P. (2019).
\newblock Adversarial uncertainty quantification in physics-informed neural
  networks.
\newblock {\em Journal of Computational Physics}, 394:136--152.

\bibitem[Yeh et~al., 2016]{yeh2016semantic}
Yeh, R., Chen, C., Lim, T.~Y., Hasegawa-Johnson, M., and Do, M.~N. (2016).
\newblock Semantic image inpainting with perceptual and contextual losses.
\newblock {\em arXiv preprint arXiv:1607.07539}, 2(3).

\bibitem[Zhang et~al., 2019]{Zhang2019}
Zhang, K., Zhong, G., Dong, J., Wang, S., and Wang, Y. (2019).
\newblock {Stock market prediction based on generative adversarial network}.
\newblock {\em Procedia Computer Science}, 147:400--406.

\bibitem[Zhu et~al., 2020]{zhu2020deconstructing}
Zhu, B., Jiao, J., and Tse, D. (2020).
\newblock Deconstructing generative adversarial networks.
\newblock {\em IEEE Transactions on Information Theory}.

\bibitem[Zhu et~al., 2016]{zhu2016generative}
Zhu, J.-Y., Kr{\"a}henb{\"u}hl, P., Shechtman, E., and Efros, A.~A. (2016).
\newblock Generative visual manipulation on the natural image manifold.
\newblock In {\em European Conference on Computer Vision}, pages 597--613.
  Springer.

\end{thebibliography}

\stopsubchapters
\end{document}